\documentclass[a4paper,twocolumn,11pt,accepted=2023-10-10]{quantumarticle}
\pdfoutput=1
\usepackage{amsmath}
\usepackage{pifont}
\usepackage{amsthm}
\usepackage{qcircuit}
\usepackage{cancel}
\usepackage{changes,subcaption,comment}
\usepackage[numbers,compress]{natbib}

\def\Hy{{\sf H}}
\def\A{{\sf A}}
\def\B{{\sf B}}
\def\sC{{\sf C}}
\def\G{{\sf G}}
\def\S{{\sf S}}

\def\Q{{\sf Q}}

\def\mS{\Omega}

\def\Tr{{\rm Tr}}

\def\C{{\mathbb C}}
\def\Hi{{\mathcal H}}

\def\mI{{\mathcal I}}

\def\E{{\mathcal E}}

\def\T{{\mathcal T}}
\def\R{{\mathtt R}}
\def\m{{\mathtt m}}
\def\P{{\mathtt P}}
\def\f{{\mathtt f}}
\def\s{{\mathtt s}}

\def\I{{\mathbb I}}

\usepackage{rotating}
\usepackage{floatpag}
\rotfloatpagestyle{empty}
\renewcommand{\theenumi}{\arabic{enumi}}
\usepackage{hyperref}
\usepackage{dsfont}
\usepackage{amsmath,amsthm,amssymb}
\usepackage{xspace,enumerate,epsfig}
\usepackage{graphicx}
\graphicspath{{.}{./figures/}}
\usepackage{marvosym}

\usepackage{tikzfig}
\usepackage{stmaryrd}
\usepackage{docmute}
\usepackage{keycommand}

\newkeycommand{\pointmap}[style=point][1]{\,\tikz{\node[style=\commandkey{style}] (x) at (0,-0.3) {$#1$};\node [style=none] (3) at (0, 1.0) {};\node [style=none] (3) at (0, -1.0) {};\draw (x)--(0,0.7);}\,}

\newkeycommand{\typedkpoint}[style=kpoint,edgestyle=][2]{%
\,\begin{tikzpicture}
  \begin{pgfonlayer}{nodelayer}
    \node [style=none] (0) at (0, 1) {};
    \node [style=\commandkey{style}] (1) at (0, -0.75) {$#2$};
    \node [style=right label] (2) at (0.25, 0.5) {$#1$};
  \end{pgfonlayer}
  \begin{pgfonlayer}{edgelayer}
    \draw [\commandkey{edgestyle}] (1) to (0.center);
  \end{pgfonlayer}
\end{tikzpicture}\,}

\newkeycommand{\typedkpointdag}[style=kpoint adjoint,edgestyle=][2]{%
\,\begin{tikzpicture}
  \begin{pgfonlayer}{nodelayer}
    \node [style=none] (0) at (0, -1) {};
    \node [style=\commandkey{style}] (1) at (0, 0.75) {$#2$};
    \node [style=right label] (2) at (0.25, -0.5) {$#1$};
  \end{pgfonlayer}
  \begin{pgfonlayer}{edgelayer}
    \draw [\commandkey{edgestyle}] (0.center) to (1);
  \end{pgfonlayer}
\end{tikzpicture}\,}

\newcommand{\bistateadj}[1]{\,\begin{tikzpicture}[yshift=-3mm]
  \begin{pgfonlayer}{nodelayer}
    \node [style=kpointadj, minimum width=1 cm, inner sep=2pt] (0) at (0, 0.5) {$\psi$};
    \node [style=none] (1) at (-0.75, 0.25) {};
    \node [style=none] (2) at (0.75, 0.25) {};
    \node [style=none] (3) at (-0.75, -0.5) {};
    \node [style=none] (4) at (0.75, -0.5) {};
  \end{pgfonlayer}
  \begin{pgfonlayer}{edgelayer}
    \draw (1.center) to (3.center);
    \draw (2.center) to (4.center);
  \end{pgfonlayer}
\end{tikzpicture}\,}

\newkeycommand{\pointketbra}[style1=copoint,style2=point][2]{\,%
\begin{tikzpicture}
  \begin{pgfonlayer}{nodelayer}
    \node [style=none] (0) at (0, -1.5) {};
    \node [style=\commandkey{style1}] (1) at (0, -0.75) {$#1$};
    \node [style=\commandkey{style2}] (2) at (0, 0.75) {$#2$};
    \node [style=none] (3) at (0, 1.5) {};
  \end{pgfonlayer}
  \begin{pgfonlayer}{edgelayer}
    \draw (0.center) to (1);
    \draw (2) to (3.center);
  \end{pgfonlayer}
\end{tikzpicture}\,}

\newkeycommand{\twocopointketbra}[style1=copoint,style2=copoint,style3=point][3]{\,%
\begin{tikzpicture}
  \begin{pgfonlayer}{nodelayer}
    \node [style=none] (0) at (-0.75, -1.5) {};
    \node [style=none] (1) at (0.75, -1.5) {};
    \node [style=\commandkey{style1}] (2) at (-0.75, -0.75) {$#1$};
    \node [style=\commandkey{style2}] (3) at (0.75, -0.75) {$#2$};
    \node [style=\commandkey{style3}] (4) at (0, 0.75) {$#3$};
    \node [style=none] (5) at (0, 1.5) {};
  \end{pgfonlayer}
  \begin{pgfonlayer}{edgelayer}
    \draw (0.center) to (2);
    \draw (1.center) to (3);
    \draw (4) to (5.center);
  \end{pgfonlayer}
\end{tikzpicture}\,}

\newkeycommand{\twopointketbra}[style1=copoint,style2=point,style3=point][3]{\,%
\begin{tikzpicture}
  \begin{pgfonlayer}{nodelayer}
    \node [style=none] (0) at (0, -1.5) {};
    \node [style=none] (1) at (-0.75, 1.5) {};
    \node [style=\commandkey{style1}] (2) at (0, -0.75) {$#1$};
    \node [style=\commandkey{style2}] (3) at (-0.75, 0.75) {$#2$};
    \node [style=\commandkey{style3}] (4) at (0.75, 0.75) {$#3$};
    \node [style=none] (5) at (0.75, 1.5) {};
  \end{pgfonlayer}
  \begin{pgfonlayer}{edgelayer}
    \draw (0.center) to (2);
    \draw (1.center) to (3);
    \draw (4) to (5.center);
  \end{pgfonlayer}
\end{tikzpicture}\,}

\newkeycommand{\pointbraket}[style1=point,style2=copoint][2]{\,%
\begin{tikzpicture}
  \begin{pgfonlayer}{nodelayer}
    \node [style=\commandkey{style1}] (0) at (0, -0.5) {$#1$};
    \node [style=\commandkey{style2}] (1) at (0, 0.5) {$#2$};
  \end{pgfonlayer}
  \begin{pgfonlayer}{edgelayer}
    \draw (0) to (1);
  \end{pgfonlayer}
\end{tikzpicture}\,}

\newkeycommand{\kpointbraket}[style1=kpoint,style2=kpoint adjoint][2]{\,%
\begin{tikzpicture}
  \begin{pgfonlayer}{nodelayer}
    \node [style=\commandkey{style1}] (0) at (0, -0.75) {$#1$};
    \node [style=\commandkey{style2}] (1) at (0, 0.75) {$#2$};
  \end{pgfonlayer}
  \begin{pgfonlayer}{edgelayer}
    \draw (0) to (1);
  \end{pgfonlayer}
\end{tikzpicture}\,}

\newkeycommand{\kpointmap}[style1=kpoint,style2=map][2]{\,%
\begin{tikzpicture}
  \begin{pgfonlayer}{nodelayer}
    \node [style=\commandkey{style1}] (0) at (0, -1.1) {$#1$};
    \node [style=\commandkey{style2}] (1) at (0, 0.2) {$#2$};
    \node [style=none] (2) at (0, 1.5) {};
  \end{pgfonlayer}
  \begin{pgfonlayer}{edgelayer}
    \draw (0) to (1);
    \draw (1) to (2.center);
  \end{pgfonlayer}
\end{tikzpicture}%
\,}

\newkeycommand{\sandwichtwo}[style1=point,style2=map,style3=map,style4=copoint][4]{\,%
\begin{tikzpicture}
  \begin{pgfonlayer}{nodelayer}
    \node [style=\commandkey{style2}] (0) at (0, -0.75) {$#2$};
    \node [style=\commandkey{style3}] (1) at (0, 0.75) {$#3$};
    \node [style=\commandkey{style1}] (2) at (0, -2) {$#1$};
    \node [style=\commandkey{style4}] (3) at (0, 2) {$#4$};
  \end{pgfonlayer}
  \begin{pgfonlayer}{edgelayer}
    \draw (2) to (0);
    \draw (0) to (1);
    \draw (1) to (3);
  \end{pgfonlayer}
\end{tikzpicture}\,}

\newkeycommand{\longbraket}[style1=point,style2=copoint][2]{\,%
\begin{tikzpicture}
  \begin{pgfonlayer}{nodelayer}
    \node [style=\commandkey{style1}] (0) at (0, -2) {$#1$};
    \node [style=\commandkey{style2}] (1) at (0, 2) {$#2$};
  \end{pgfonlayer}
  \begin{pgfonlayer}{edgelayer}
    \draw (0) to (1);
  \end{pgfonlayer}
\end{tikzpicture}\,}

\newkeycommand{\onb}[style=point]{\ensuremath{\left\{\tikz{\node[style=\commandkey{style}] (x) at (0,-0.3) {$j$};\draw (x)--(0,0.7);}\right\}}\xspace}

\newkeycommand{\copointmap}[style=copoint][1]{\,\tikz{\node[style=\commandkey{style}] (x) at (0,0.3) {$#1$};\draw (x)--(0,-0.7);}\,}

\newcommand{\bra}[1]{\ensuremath{\left\langle #1 \right|}}
\newcommand{\ket}[1]{\ensuremath{\left|  #1 \right\rangle}}

\newcommand{\ketbra}[2]{\ensuremath{\ket{#1}\!\bra{#2}}}

\tikzstyle{cdiag}=[matrix of math nodes, row sep=3em, column sep=3em, text height=1.5ex, text depth=0.25ex,inner sep=0.5em]
\tikzstyle{arrow above}=[transform canvas={yshift=0.5ex}]
\tikzstyle{arrow below}=[transform canvas={yshift=-0.5ex}]

\usetikzlibrary{shapes.multipart}

\tikzstyle{bwSpider}=[
       rectangle split,
       rectangle split parts=2,
       rectangle split part fill={black,white},
 minimum size=3.6 mm, inner sep=-2mm, draw=black,scale=0.5,rounded corners=0.8 mm
       ]
 \tikzstyle{wbSpider}=[
       rectangle split,
       rectangle split parts=2,
       rectangle split part fill={white,black},
 minimum size=3.6 mm, inner sep=-2mm, draw=black,scale=0.5,rounded corners=0.8 mm
       ]
\tikzstyle{qWire}=[line width = 1pt, color=black]
\tikzstyle{cWire}=[color=gray,thin]
\tikzstyle{env}=[copoint,regular polygon rotate=0,minimum width=0.2cm, fill=black]

\tikzstyle{probs}=[shape=semicircle,fill=white,draw=black,shape border rotate=180,minimum width=1.2cm]

\tikzstyle{every picture}=[baseline=-0.25em,scale=0.5]
\tikzstyle{dotpic}=[] 
\tikzstyle{diredges}=[every to/.style={diredge}]
\tikzstyle{math matrix}=[matrix of math nodes,left delimiter=(,right delimiter=),inner sep=2pt,column sep=1em,row sep=0.5em,nodes={inner sep=0pt},text height=1.5ex, text depth=0.25ex]

\tikzstyle{inline text}=[text height=1.5ex, text depth=0.25ex,yshift=0.5mm]
\tikzstyle{label}=[font=\footnotesize,text height=1.5ex, text depth=0.25ex,yshift=0.5mm]
\tikzstyle{left label}=[label,anchor=east,xshift=1.5mm]
\tikzstyle{right label}=[label,anchor=west,xshift=-1mm]

\tikzstyle{braceedge}=[decorate,decoration={brace,amplitude=2mm,raise=-1mm}]
\tikzstyle{small braceedge}=[decorate,decoration={brace,amplitude=1mm,raise=-1mm}]

\tikzstyle{doubled}=[line width=1.6pt] 
\tikzstyle{boldedge}=[doubled,shorten <=-0.17mm,shorten >=-0.17mm]
\tikzstyle{boldedgegray}=[doubled,gray,shorten <=-0.17mm,shorten >=-0.17mm]
\tikzstyle{singleedgegray}=[gray]

\tikzstyle{semidoubled}=[line width=1.4pt] 
\tikzstyle{semiboldedgegray}=[semidoubled,gray,shorten <=-0.17mm,shorten >=-0.17mm]

\tikzstyle{boxedge}=[semiboldedgegray]

\tikzstyle{boldedgedashed}=[very thick,dashed,shorten <=-0.17mm,shorten >=-0.17mm]
\tikzstyle{vboldedgedashed}=[doubled,dashed,shorten <=-0.17mm,shorten >=-0.17mm]
\tikzstyle{left hook arrow}=[left hook-latex]
\tikzstyle{right hook arrow}=[right hook-latex]
\tikzstyle{sembracket}=[line width=0.5pt,shorten <=-0.07mm,shorten >=-0.07mm]

\tikzstyle{causal edge}=[->,thick,gray]
\tikzstyle{causal nondir}=[thick,gray]
\tikzstyle{timeline}=[thick,gray, dashed]

\tikzstyle{cedge}=[<->,thick,gray!70!white]

\tikzstyle{empty diagram}=[draw=gray!40!white,dashed,shape=rectangle,minimum width=1cm,minimum height=1cm]
\tikzstyle{empty diagram small}=[draw=gray!50!white,dashed,shape=rectangle,minimum width=0.6cm,minimum height=0.5cm]

\tikzstyle{dot}=[inner sep=0mm,minimum width=2mm,minimum height=2mm,draw,shape=circle]
\tikzstyle{leak}=[white dot, shape=regular polygon, minimum size=3.3 mm, regular polygon sides=3, outer sep=-0.2mm, regular polygon rotate=270]
\tikzstyle{proj}=[draw,fill=white,chamfered rectangle,chamfered rectangle angle=30, minimum width=2mm, minimum height= 1mm,scale=0.5, outer sep=-0.2mm]
\tikzstyle{wide proj}=[draw,fill=white,chamfered rectangle,chamfered rectangle angle=30, minimum width=15mm,minimum height=1mm,scale=0.5, outer sep=-0.2mm]
\tikzstyle{very wide proj}=[draw,fill=white,chamfered rectangle,chamfered rectangle angle=30, minimum width=25mm,minimum height=1mm,scale=0.5, outer sep=-0.2mm]
\tikzstyle{very very wide proj}=[draw,fill=white,chamfered rectangle,chamfered rectangle angle=30, minimum width=35mm,minimum height=1mm,scale=0.5, outer sep=-0.2mm]
\tikzstyle{preleak}=[proj]

\tikzstyle{split proj out}=[regular polygon,regular polygon sides=3,draw,scale=0.75,inner sep=-0.5pt,minimum width=3.3mm,fill=white,regular polygon rotate=180]
\tikzstyle{split proj in}=[regular polygon,regular polygon sides=3,draw,scale=0.75,inner sep=-0.5pt,minimum width=3.3mm,fill=white]
\tikzstyle{Vleak}=[white dot, shape=regular polygon, minimum size=3.3 mm, regular polygon sides=3, outer sep=-0.2mm, regular polygon rotate=90]
\tikzstyle{dleak}=[white dot, line width=1.6pt, shape=regular polygon, minimum size=3.3 mm, regular polygon sides=3, outer sep=-0.2mm, regular polygon rotate=270]

\tikzstyle{Wsquare}=[white dot, shape=regular polygon, rounded corners=0.8 mm, minimum size=3.3 mm, regular polygon sides=3, outer sep=-0.2mm]
\tikzstyle{Wsquareadj}=[white dot, shape=regular polygon, rounded corners=0.8 mm, minimum size=3.3 mm, regular polygon sides=3, outer sep=-0.2mm, regular polygon rotate=180]
\tikzstyle{ddot}=[inner sep=0mm, doubled, minimum width=2.5mm,minimum height=2.5mm,draw,shape=circle]

\tikzstyle{black dot}=[dot,fill=black]
\tikzstyle{white dot}=[dot,fill=white,,text depth=-0.2mm]
\tikzstyle{white Wsquare}=[Wsquare,fill=gray,,text depth=-0.2mm]
\tikzstyle{white Wsquareadj}=[Wsquareadj,fill=white,,text depth=-0.2mm]
\tikzstyle{green dot}=[white dot] 
\tikzstyle{gray dot}=[dot,fill=gray!40!white,,text depth=-0.2mm]
\tikzstyle{red dot}=[gray dot] 

\tikzstyle{black ddot}=[ddot,fill=black]
\tikzstyle{white ddot}=[ddot,fill=white]
\tikzstyle{gray ddot}=[ddot,fill=gray!40!white]

\tikzstyle{gray edge}=[gray!60!white]

\tikzstyle{small dot}=[inner sep=0.5mm,minimum width=0pt,minimum height=0pt,draw,shape=circle]

\tikzstyle{small black dot}=[small dot,fill=black]
\tikzstyle{small white dot}=[small dot,fill=white]
\tikzstyle{small gray dot}=[small dot,fill=gray!40!white]

\tikzstyle{causal dot}=[inner sep=0.4mm,minimum width=0pt,minimum height=0pt,draw=white,shape=circle,fill=gray!40!white]

\tikzstyle{phase dimensions}=[minimum size=5mm,font=\footnotesize,rectangle,rounded corners=2.5mm,inner sep=0.2mm,outer sep=-2mm]
\tikzstyle{dphase dimensions}=[minimum size=5mm,font=\footnotesize,rectangle,rounded corners=2.5mm,inner sep=0.2mm,outer sep=-2mm]

\tikzstyle{white phase dot}=[dot,fill=white,phase dimensions]
\tikzstyle{white phase ddot}=[ddot,fill=white,dphase dimensions]

\tikzstyle{white rect ddot}=[draw=black,fill=white,doubled,minimum size=5mm,font=\footnotesize,rectangle,rounded corners=2.5mm,inner sep=0.2mm]
\tikzstyle{gray rect ddot}=[draw=black,fill=gray!40!white,doubled,minimum size=6mm,font=\footnotesize,rectangle,rounded corners=3mm]

\tikzstyle{gray phase dot}=[dot,fill=gray!40!white,phase dimensions]
\tikzstyle{gray phase ddot}=[ddot,fill=gray!40!white,dphase dimensions]
\tikzstyle{grey phase dot}=[gray phase dot]
\tikzstyle{grey phase ddot}=[gray phase ddot]

\tikzstyle{small phase dimensions}=[minimum size=4mm,font=\tiny,rectangle,rounded corners=2mm,inner sep=0.2mm,outer sep=-2mm]
\tikzstyle{small dphase dimensions}=[minimum size=4mm,font=\tiny,rectangle,rounded corners=2mm,inner sep=0.2mm,outer sep=-2mm]

\tikzstyle{small gray phase dot}=[dot,fill=gray!40!white,small phase dimensions]
\tikzstyle{small gray phase ddot}=[ddot,fill=gray!40!white,small dphase dimensions]

\tikzstyle{small map}=[draw,shape=rectangle,minimum height=4mm,minimum width=4mm,fill=white]

\tikzstyle{cnot}=[fill=white,shape=circle,inner sep=-1.4pt]

\tikzstyle{asym hadamard}=[fill=white,draw,shape=NEbox,inner sep=0.6mm,font=\footnotesize,minimum height=4mm]
\tikzstyle{asym hadamard conj}=[fill=white,draw,shape=NWbox,inner sep=0.6mm,font=\footnotesize,minimum height=4mm]
\tikzstyle{asym hadamard dag}=[fill=white,draw,shape=SEbox,inner sep=0.6mm,font=\footnotesize,minimum height=4mm]

\tikzstyle{hadamard}=[fill=white,draw,inner sep=0.6mm,font=\footnotesize,minimum height=4mm,minimum width=4mm]
\tikzstyle{small hadamard}=[fill=white,draw,inner sep=0.6mm,minimum height=1.5mm,minimum width=1.5mm]
\tikzstyle{small hadamard rotate}=[small hadamard,rotate=45]
\tikzstyle{dhadamard}=[hadamard,doubled]
\tikzstyle{small dhadamard}=[small hadamard,doubled]
\tikzstyle{small dhadamard rotate}=[small hadamard rotate,doubled]
\tikzstyle{antipode}=[white dot,inner sep=0.3mm,font=\footnotesize]

\tikzstyle{scalar}=[diamond,draw,inner sep=0.5pt,font=\small]
\tikzstyle{dscalar}=[diamond,doubled, draw,inner sep=0.5pt,font=\small]

\tikzstyle{small box}=[rectangle,inline text,fill=white,draw,minimum height=5mm,yshift=-0.5mm,minimum width=5mm,font=\small]
\tikzstyle{small gray box}=[small box,fill=gray!30]
\tikzstyle{medium box}=[rectangle,inline text,fill=white,draw,minimum height=5mm,yshift=-0.5mm,minimum width=10mm,font=\small]
\tikzstyle{square box}=[small box] 
\tikzstyle{medium gray box}=[small box,fill=gray!30]
\tikzstyle{semilarge box}=[rectangle,inline text,fill=white,draw,minimum height=5mm,yshift=-0.5mm,minimum width=12.5mm,font=\small]
\tikzstyle{large box}=[rectangle,inline text,fill=white,draw,minimum height=5mm,yshift=-0.5mm,minimum width=15mm,font=\small]
\tikzstyle{large gray box}=[small box,fill=gray!30]

\tikzstyle{Bayes box}=[rectangle,fill=black,draw, minimum height=3mm, minimum width=3mm]

\tikzstyle{gray square point}=[small box,fill=gray!50]

\tikzstyle{dphase box white}=[dhadamard]
\tikzstyle{dphase box gray}=[dhadamard,fill=gray!50!white]
\tikzstyle{phase box white}=[hadamard]
\tikzstyle{phase box gray}=[hadamard,fill=gray!50!white]

\tikzstyle{point}=[regular polygon,regular polygon sides=3,draw,scale=0.75,inner sep=-0.5pt,minimum width=9mm,fill=white,regular polygon rotate=180]
\tikzstyle{point nosep}=[regular polygon,regular polygon sides=3,draw,scale=0.75,inner sep=-2pt,minimum width=9mm,fill=white,regular polygon rotate=180]
\tikzstyle{copoint}=[regular polygon,regular polygon sides=3,draw,scale=0.75,inner sep=-0.5pt,minimum width=9mm,fill=white]
\tikzstyle{dpoint}=[point,doubled]
\tikzstyle{dcopoint}=[copoint,doubled]

\tikzstyle{pointgrow}=[shape=cornerpoint,kpoint common,scale=0.75,inner sep=3pt]
\tikzstyle{pointgrow dag}=[shape=cornercopoint,kpoint common,scale=0.75,inner sep=3pt]

\tikzstyle{wide copoint}=[fill=white,draw,shape=isosceles triangle,shape border rotate=90,isosceles triangle stretches=true,inner sep=0pt,minimum width=1.5cm,minimum height=6.12mm]
\tikzstyle{wide point}=[fill=white,draw,shape=isosceles triangle,shape border rotate=-90,isosceles triangle stretches=true,inner sep=0pt,minimum width=1.5cm,minimum height=6.12mm,yshift=-0.0mm]
\tikzstyle{wide point plus}=[fill=white,draw,shape=isosceles triangle,shape border rotate=-90,isosceles triangle stretches=true,inner sep=0pt,minimum width=1.74cm,minimum height=7mm,yshift=-0.0mm]

\tikzstyle{wide dpoint}=[fill=white,doubled,draw,shape=isosceles triangle,shape border rotate=-90,isosceles triangle stretches=true,inner sep=0pt,minimum width=1.5cm,minimum height=6.12mm,yshift=-0.0mm]

\tikzstyle{tinypoint}=[regular polygon,regular polygon sides=3,draw,scale=0.55,inner sep=-0.15pt,minimum width=6mm,fill=white,regular polygon rotate=180]

\tikzstyle{white point}=[point]
\tikzstyle{white dpoint}=[dpoint]
\tikzstyle{green point}=[white point] 
\tikzstyle{white copoint}=[copoint]
\tikzstyle{gray point}=[point,fill=gray!40!white]
\tikzstyle{gray dpoint}=[gray point,doubled]
\tikzstyle{red point}=[gray point] 
\tikzstyle{gray copoint}=[copoint,fill=gray!40!white]
\tikzstyle{gray dcopoint}=[gray copoint,doubled]

\tikzstyle{white point guide}=[regular polygon,regular polygon sides=3,font=\scriptsize,draw,scale=0.65,inner sep=-0.5pt,minimum width=9mm,fill=white,regular polygon rotate=180]

\tikzstyle{black point}=[point,fill=black,font=\color{white}]
\tikzstyle{black copoint}=[copoint,fill=black,font=\color{white}]

\tikzstyle{tiny gray point}=[tinypoint,fill=gray!40!white]

\tikzstyle{diredge}=[->]
\tikzstyle{ddiredge}=[<->]
\tikzstyle{rdiredge}=[<-]
\tikzstyle{thickdiredge}=[->, very thick]
\tikzstyle{pointer edge}=[->,very thick,gray]
\tikzstyle{pointer edge part}=[very thick,gray]
\tikzstyle{dashed edge}=[dashed]
\tikzstyle{thick dashed edge}=[very thick,dashed]
\tikzstyle{thick gray dashed edge}=[thick dashed edge,gray!40]
\tikzstyle{thick map edge}=[very thick,|->]

\makeatletter
\newcommand{\boxshape}[3]{%
\pgfdeclareshape{#1}{
\inheritsavedanchors[from=rectangle] 
\inheritanchorborder[from=rectangle]
\inheritanchor[from=rectangle]{center}
\inheritanchor[from=rectangle]{north}
\inheritanchor[from=rectangle]{south}
\inheritanchor[from=rectangle]{west}
\inheritanchor[from=rectangle]{east}
\backgroundpath{
\southwest \pgf@xa=\pgf@x \pgf@ya=\pgf@y
\northeast \pgf@xb=\pgf@x \pgf@yb=\pgf@y

\@tempdima=#2
\@tempdimb=#3

\pgfpathmoveto{\pgfpoint{\pgf@xa - 5pt + \@tempdima}{\pgf@ya}}
\pgfpathlineto{\pgfpoint{\pgf@xa - 5pt - \@tempdima}{\pgf@yb}}
\pgfpathlineto{\pgfpoint{\pgf@xb + 5pt + \@tempdimb}{\pgf@yb}}
\pgfpathlineto{\pgfpoint{\pgf@xb + 5pt - \@tempdimb}{\pgf@ya}}
\pgfpathlineto{\pgfpoint{\pgf@xa - 5pt + \@tempdima}{\pgf@ya}}
\pgfpathclose
}
}}

\boxshape{NEbox}{0pt}{3pt}
\boxshape{SEbox}{0pt}{-3pt}
\boxshape{NWbox}{5pt}{0pt}
\boxshape{SWbox}{-5pt}{0pt}
\boxshape{EBox}{-3pt}{3pt}
\boxshape{WBox}{3pt}{-3pt}
\makeatother

\tikzstyle{cloud}=[shape=cloud,draw,minimum width=1.5cm,minimum height=1.5cm]

\tikzstyle{map}=[draw,shape=NEbox,inner sep=1pt,minimum height=4mm,fill=white]
\tikzstyle{dashedmap}=[draw,dashed,shape=NEbox,inner sep=2pt,minimum height=6mm,fill=white]
\tikzstyle{mapdag}=[draw,shape=SEbox,inner sep=1pt,minimum height=4mm,fill=white]
\tikzstyle{mapadj}=[draw,shape=SEbox,inner sep=2pt,minimum height=6mm,fill=white]
\tikzstyle{maptrans}=[draw,shape=SWbox,inner sep=2pt,minimum height=6mm,fill=white]
\tikzstyle{mapconj}=[draw,shape=NWbox,inner sep=2pt,minimum height=6mm,fill=white]

\tikzstyle{medium map}=[draw,shape=NEbox,inner sep=2pt,minimum height=6mm,fill=white,minimum width=7mm]
\tikzstyle{medium map dag}=[draw,shape=SEbox,inner sep=2pt,minimum height=6mm,fill=white,minimum width=7mm]
\tikzstyle{medium map adj}=[draw,shape=SEbox,inner sep=2pt,minimum height=6mm,fill=white,minimum width=7mm]
\tikzstyle{medium map trans}=[draw,shape=SWbox,inner sep=2pt,minimum height=6mm,fill=white,minimum width=7mm]
\tikzstyle{medium map conj}=[draw,shape=NWbox,inner sep=2pt,minimum height=6mm,fill=white,minimum width=7mm]
\tikzstyle{semilarge map}=[draw,shape=NEbox,inner sep=2pt,minimum height=6mm,fill=white,minimum width=9.5mm]
\tikzstyle{semilarge map trans}=[draw,shape=SWbox,inner sep=2pt,minimum height=6mm,fill=white,minimum width=9.5mm]
\tikzstyle{semilarge map adj}=[draw,shape=SEbox,inner sep=2pt,minimum height=6mm,fill=white,minimum width=9.5mm]
\tikzstyle{semilarge map dag}=[draw,shape=SEbox,inner sep=2pt,minimum height=6mm,fill=white,minimum width=9.5mm]
\tikzstyle{semilarge map conj}=[draw,shape=NWbox,inner sep=2pt,minimum height=6mm,fill=white,minimum width=9.5mm]
\tikzstyle{large map}=[draw,shape=NEbox,inner sep=2pt,minimum height=6mm,fill=white,minimum width=12mm]
\tikzstyle{large map conj}=[draw,shape=NWbox,inner sep=2pt,minimum height=6mm,fill=white,minimum width=12mm]
\tikzstyle{very large map}=[draw,shape=NEbox,inner sep=2pt,minimum height=6mm,fill=white,minimum width=17mm]

\tikzstyle{medium dmap}=[draw,doubled,shape=NEbox,inner sep=2pt,minimum height=6mm,fill=white,minimum width=7mm]
\tikzstyle{medium dmap dag}=[draw,doubled,shape=SEbox,inner sep=2pt,minimum height=6mm,fill=white,minimum width=7mm]
\tikzstyle{medium dmap adj}=[draw,doubled,shape=SEbox,inner sep=2pt,minimum height=6mm,fill=white,minimum width=7mm]
\tikzstyle{medium dmap trans}=[draw,doubled,shape=SWbox,inner sep=2pt,minimum height=6mm,fill=white,minimum width=7mm]
\tikzstyle{medium dmap conj}=[draw,doubled,shape=NWbox,inner sep=2pt,minimum height=6mm,fill=white,minimum width=7mm]
\tikzstyle{semilarge dmap}=[draw,doubled,shape=NEbox,inner sep=2pt,minimum height=6mm,fill=white,minimum width=9.5mm]
\tikzstyle{semilarge dmap trans}=[draw,doubled,shape=SWbox,inner sep=2pt,minimum height=6mm,fill=white,minimum width=9.5mm]
\tikzstyle{semilarge dmap adj}=[draw,doubled,shape=SEbox,inner sep=2pt,minimum height=6mm,fill=white,minimum width=9.5mm]
\tikzstyle{semilarge dmap dag}=[draw,doubled,shape=SEbox,inner sep=2pt,minimum height=6mm,fill=white,minimum width=9.5mm]
\tikzstyle{semilarge dmap conj}=[draw,doubled,shape=NWbox,inner sep=2pt,minimum height=6mm,fill=white,minimum width=9.5mm]
\tikzstyle{large dmap}=[draw,doubled,shape=NEbox,inner sep=2pt,minimum height=6mm,fill=white,minimum width=12mm]
\tikzstyle{large dmap conj}=[draw,doubled,shape=NWbox,inner sep=2pt,minimum height=6mm,fill=white,minimum width=12mm]
\tikzstyle{large dmap trans}=[draw,doubled,shape=SWbox,inner sep=2pt,minimum height=6mm,fill=white,minimum width=12mm]
\tikzstyle{large dmap adj}=[draw,doubled,shape=SEbox,inner sep=2pt,minimum height=6mm,fill=white,minimum width=12mm]
\tikzstyle{large dmap dag}=[draw,doubled,shape=SEbox,inner sep=2pt,minimum height=6mm,fill=white,minimum width=12mm]
\tikzstyle{very large dmap}=[draw,doubled,shape=NEbox,inner sep=2pt,minimum height=6mm,fill=white,minimum width=19.5mm]

\tikzstyle{muxbox}=[draw,shape=rectangle,minimum height=3mm,minimum width=3mm,fill=white]
\tikzstyle{dmuxbox}=[muxbox,doubled]

\tikzstyle{box}=[draw,shape=rectangle,inner sep=2pt,minimum height=6mm,minimum width=6mm,fill=white]
\tikzstyle{dbox}=[draw,doubled,shape=rectangle,inner sep=2pt,minimum height=6mm,minimum width=6mm,fill=white]
\tikzstyle{dmap}=[draw,doubled,shape=NEbox,inner sep=2pt,minimum height=6mm,fill=white]
\tikzstyle{dmapdag}=[draw,doubled,shape=SEbox,inner sep=2pt,minimum height=6mm,fill=white]
\tikzstyle{dmapadj}=[draw,doubled,shape=SEbox,inner sep=2pt,minimum height=6mm,fill=white]
\tikzstyle{dmaptrans}=[draw,doubled,shape=SWbox,inner sep=2pt,minimum height=6mm,fill=white]
\tikzstyle{dmapconj}=[draw,doubled,shape=NWbox,inner sep=2pt,minimum height=6mm,fill=white]

\tikzstyle{ddmap}=[draw,doubled,dashed,shape=NEbox,inner sep=2pt,minimum height=6mm,fill=white]
\tikzstyle{ddmapdag}=[draw,doubled,dashed,shape=SEbox,inner sep=2pt,minimum height=6mm,fill=white]
\tikzstyle{ddmapadj}=[draw,doubled,dashed,shape=SEbox,inner sep=2pt,minimum height=6mm,fill=white]
\tikzstyle{ddmaptrans}=[draw,doubled,dashed,shape=SWbox,inner sep=2pt,minimum height=6mm,fill=white]
\tikzstyle{ddmapconj}=[draw,doubled,dashed,shape=NWbox,inner sep=2pt,minimum height=6mm,fill=white]

\boxshape{sNEbox}{0pt}{3pt}
\boxshape{sSEbox}{0pt}{-3pt}
\boxshape{sNWbox}{3pt}{0pt}
\boxshape{sSWbox}{-3pt}{0pt}
\tikzstyle{smap}=[draw,shape=sNEbox,fill=white]
\tikzstyle{smapdag}=[draw,shape=sSEbox,fill=white]
\tikzstyle{smapadj}=[draw,shape=sSEbox,fill=white]
\tikzstyle{smaptrans}=[draw,shape=sSWbox,fill=white]
\tikzstyle{smapconj}=[draw,shape=sNWbox,fill=white]

\tikzstyle{dsmap}=[draw,dashed,shape=sNEbox,fill=white]
\tikzstyle{dsmapdag}=[draw,dashed,shape=sSEbox,fill=white]
\tikzstyle{dsmaptrans}=[draw,dashed,shape=sSWbox,fill=white]
\tikzstyle{dsmapconj}=[draw,dashed,shape=sNWbox,fill=white]

\boxshape{mNEbox}{0pt}{10pt}
\boxshape{mSEbox}{0pt}{-10pt}
\boxshape{mNWbox}{10pt}{0pt}
\boxshape{mSWbox}{-10pt}{0pt}
\tikzstyle{mmap}=[draw,shape=mNEbox]
\tikzstyle{mmapdag}=[draw,shape=mSEbox]
\tikzstyle{mmaptrans}=[draw,shape=mSWbox]
\tikzstyle{mmapconj}=[draw,shape=mNWbox]

\tikzstyle{mmapgray}=[draw,fill=gray!40!white,shape=mNEbox]
\tikzstyle{smapgray}=[draw,fill=gray!40!white,shape=sNEbox]

\makeatletter

\pgfdeclareshape{cornerpoint}{
\inheritsavedanchors[from=rectangle] 
\inheritanchorborder[from=rectangle]
\inheritanchor[from=rectangle]{center}
\inheritanchor[from=rectangle]{north}
\inheritanchor[from=rectangle]{south}
\inheritanchor[from=rectangle]{west}
\inheritanchor[from=rectangle]{east}
\backgroundpath{
\southwest \pgf@xa=\pgf@x \pgf@ya=\pgf@y
\northeast \pgf@xb=\pgf@x \pgf@yb=\pgf@y

\pgfmathsetmacro{\pgf@shorten@left}{\pgfkeysvalueof{/tikz/shorten left}}
\pgfmathsetmacro{\pgf@shorten@right}{\pgfkeysvalueof{/tikz/shorten right}}

\pgfpathmoveto{\pgfpoint{0.5 * (\pgf@xa + \pgf@xb)}{\pgf@ya - 5pt}}
\pgfpathlineto{\pgfpoint{\pgf@xa - 8pt + \pgf@shorten@left}{\pgf@yb - 1.5 * \pgf@shorten@left}}
\pgfpathlineto{\pgfpoint{\pgf@xa - 8pt + \pgf@shorten@left}{\pgf@yb}}
\pgfpathlineto{\pgfpoint{\pgf@xb + 8pt - \pgf@shorten@right}{\pgf@yb}}
\pgfpathlineto{\pgfpoint{\pgf@xb + 8pt - \pgf@shorten@right}{\pgf@yb - 1.5 * \pgf@shorten@right}}
\pgfpathclose
}
}

\pgfdeclareshape{cornercopoint}{
\inheritsavedanchors[from=rectangle] 
\inheritanchorborder[from=rectangle]
\inheritanchor[from=rectangle]{center}
\inheritanchor[from=rectangle]{north}
\inheritanchor[from=rectangle]{south}
\inheritanchor[from=rectangle]{west}
\inheritanchor[from=rectangle]{east}
\backgroundpath{
\southwest \pgf@xa=\pgf@x \pgf@ya=\pgf@y
\northeast \pgf@xb=\pgf@x \pgf@yb=\pgf@y

\pgfmathsetmacro{\pgf@shorten@left}{\pgfkeysvalueof{/tikz/shorten left}}
\pgfmathsetmacro{\pgf@shorten@right}{\pgfkeysvalueof{/tikz/shorten right}}

\pgfpathmoveto{\pgfpoint{0.5 * (\pgf@xa + \pgf@xb)}{\pgf@yb + 5pt}}
\pgfpathlineto{\pgfpoint{\pgf@xa - 8pt + \pgf@shorten@left}{\pgf@ya + 1.5 * \pgf@shorten@left}}
\pgfpathlineto{\pgfpoint{\pgf@xa - 8pt + \pgf@shorten@left}{\pgf@ya}}
\pgfpathlineto{\pgfpoint{\pgf@xb + 8pt - \pgf@shorten@right}{\pgf@ya}}
\pgfpathlineto{\pgfpoint{\pgf@xb + 8pt - \pgf@shorten@right}{\pgf@ya + 1.5 * \pgf@shorten@right}}
\pgfpathclose
}
}

\makeatother

\pgfkeyssetvalue{/tikz/shorten left}{0pt}
\pgfkeyssetvalue{/tikz/shorten right}{0pt}

\tikzstyle{kpoint common}=[draw,fill=white,inner sep=1pt,minimum height=4mm]
\tikzstyle{kpoint sc}=[shape=cornerpoint,kpoint common]
\tikzstyle{kpoint adjoint sc}=[shape=cornercopoint,kpoint common]
\tikzstyle{kpoint}=[shape=cornerpoint,shorten left=5pt,kpoint common]
\tikzstyle{kpoint adjoint}=[shape=cornercopoint,shorten left=5pt,kpoint common]
\tikzstyle{kpoint conjugate}=[shape=cornerpoint,shorten right=5pt,kpoint common]
\tikzstyle{kpoint transpose}=[shape=cornercopoint,shorten right=5pt,kpoint common]
\tikzstyle{kpoint symm}=[shape=cornerpoint,shorten left=5pt,shorten right=5pt,kpoint common]

\tikzstyle{wide kpoint sc}=[shape=cornerpoint,kpoint common, minimum width=1 cm]
\tikzstyle{wide kpointdag sc}=[shape=cornercopoint,kpoint common, minimum width=1 cm]

\tikzstyle{black kpoint}=[shape=cornerpoint,shorten left=5pt,kpoint common,fill=black,font=\color{white}]

\tikzstyle{black kpoint sm}=[shape=cornerpoint,shorten left=5pt,kpoint common,fill=black,font=\color{white},scale=0.75]

\tikzstyle{black kpoint adjoint}=[shape=cornercopoint,shorten left=5pt,kpoint common,fill=black,font=\color{white}]
\tikzstyle{black kpointadj}=[shape=cornercopoint,shorten left=5pt,kpoint common,fill=black,font=\color{white}]

\tikzstyle{black kpointadj sm}=[shape=cornercopoint,shorten left=5pt,kpoint common,fill=black,font=\color{white},scale=0.75]

\tikzstyle{black dkpoint}=[shape=cornerpoint,shorten left=5pt,kpoint common,fill=black, doubled,font=\color{white}]
\tikzstyle{black dkpoint adjoint}=[shape=cornercopoint,shorten left=5pt,kpoint common,fill=black, doubled,font=\color{white}]
\tikzstyle{black dkpointadj}=[shape=cornercopoint,shorten left=5pt,kpoint common,fill=black, doubled,font=\color{white}]

\tikzstyle{black dkpoint sm}=[shape=cornerpoint,shorten left=5pt,kpoint common,fill=black, doubled,font=\color{white},scale=0.75]
\tikzstyle{black dkpointadj sm}=[shape=cornercopoint,shorten left=5pt,kpoint common,fill=black, doubled,font=\color{white},scale=0.75]

\tikzstyle{kpointdag}=[kpoint adjoint]
\tikzstyle{kpointadj}=[kpoint adjoint]
\tikzstyle{kpointconj}=[kpoint conjugate]
\tikzstyle{kpointtrans}=[kpoint transpose]

\tikzstyle{big kpoint}=[kpoint, minimum width=1.2 cm, minimum height=8mm, inner sep=4pt, text depth=3mm]

\tikzstyle{wide kpoint}=[kpoint, minimum width=1 cm, inner sep=2pt]
\tikzstyle{wide kpointdag}=[kpointdag, minimum width=1 cm, inner sep=2pt]
\tikzstyle{wide kpointconj}=[kpointconj, minimum width=1 cm, inner sep=2pt]
\tikzstyle{wide kpointtrans}=[kpointtrans, minimum width=1 cm, inner sep=2pt]

\tikzstyle{wider kpoint}=[kpoint, minimum width=1.25 cm, inner sep=2pt]
\tikzstyle{wider kpointdag}=[kpointdag, minimum width=1.25 cm, inner sep=2pt]
\tikzstyle{wider kpointconj}=[kpointconj, minimum width=1.25 cm, inner sep=2pt]
\tikzstyle{wider kpointtrans}=[kpointtrans, minimum width=1.25 cm, inner sep=2pt]

\tikzstyle{gray kpoint}=[kpoint,fill=gray!50!white]
\tikzstyle{gray kpointdag}=[kpointdag,fill=gray!50!white]
\tikzstyle{gray kpointadj}=[kpointadj,fill=gray!50!white]
\tikzstyle{gray kpointconj}=[kpointconj,fill=gray!50!white]
\tikzstyle{gray kpointtrans}=[kpointtrans,fill=gray!50!white]

\tikzstyle{gray dkpoint}=[kpoint,fill=gray!50!white,doubled]
\tikzstyle{gray dkpointdag}=[kpointdag,fill=gray!50!white,doubled]
\tikzstyle{gray dkpointadj}=[kpointadj,fill=gray!50!white,doubled]
\tikzstyle{gray dkpointconj}=[kpointconj,fill=gray!50!white,doubled]
\tikzstyle{gray dkpointtrans}=[kpointtrans,fill=gray!50!white,doubled]

\tikzstyle{white label}=[draw,fill=white,rectangle,inner sep=0.7 mm]
\tikzstyle{gray label}=[draw,fill=gray!50!white,rectangle,inner sep=0.7 mm]
\tikzstyle{black label}=[draw,fill=black,rectangle,inner sep=0.7 mm]

\tikzstyle{dkpoint}=[kpoint,doubled]
\tikzstyle{wide dkpoint}=[wide kpoint,doubled]
\tikzstyle{dkpointdag}=[kpoint adjoint,doubled]
\tikzstyle{wide dkpointdag}=[wide kpointdag,doubled]
\tikzstyle{dkcopoint}=[kpoint adjoint,doubled]
\tikzstyle{dkpointadj}=[kpoint adjoint,doubled]
\tikzstyle{dkpointconj}=[kpoint conjugate,doubled]
\tikzstyle{dkpointtrans}=[kpoint transpose,doubled]

\tikzstyle{kscalar}=[kpoint common, shape=EBox, inner xsep=-1pt, inner ysep=3pt,font=\small]
\tikzstyle{kscalarconj}=[kpoint common, shape=WBox, inner xsep=-1pt, inner ysep=3pt,font=\small]

\tikzstyle{spekpoint}=[kpoint sc,minimum height=5mm,inner sep=3pt]
\tikzstyle{spekcopoint}=[kpoint adjoint sc,minimum height=5mm,inner sep=3pt]

\tikzstyle{dspekpoint}=[spekpoint,doubled]
\tikzstyle{dspekcopoint}=[spekcopoint,doubled]

 \tikzstyle{upground}=[circuit ee IEC,thick,ground,rotate=90,scale=2.5]
 \tikzstyle{downground}=[circuit ee IEC,thick,ground,rotate=-90,scale=2.5]
 \tikzstyle{bigground}=[circuit ee IEC,thick,ground,rotate=90,xscale=2.5,yscale=4.5]

\tikzstyle{arrs}=[-latex,font=\small,auto]
\tikzstyle{arrow plain}=[arrs]
\tikzstyle{arrow dashed}=[dashed,arrs]
\tikzstyle{arrow bold}=[very thick,arrs]
\tikzstyle{arrow hide}=[draw=white!0,-]
\tikzstyle{arrow reverse}=[latex-]
\tikzstyle{cdnode}=[]

\let\olddagger\dagger
\renewcommand{\dagger}{\ensuremath{\olddagger}\xspace}

\theoremstyle{definition}
\newtheorem{theorem}{Theorem}

\newtheorem{definition}[]{Definition}

\newtheorem{example}[]{Example}

\newtheorem{example*}[theorem]{Example*}
\newtheorem{examples*}[theorem]{Examples*}

\newtheorem{remark*}[theorem]{Remark*}

\theoremstyle{plain}

\usepackage{color}
\def\bR{\begin{color}{red}}
\def\bB{\begin{color}{blue}}
\def\bM{\begin{color}{magenta}}
\def\bC{\begin{color}{cyan}}
\def\bW{\begin{color}{white}}
\def\bBl{\begin{color}{black}}
\def\bG{\begin{color}{green}}
\def\bY{\begin{color}{yellow}}
\def\e{\end{color}\xspace}
\newcommand{\bit}{\begin{itemize}}
\newcommand{\eit}{\end{itemize}\par\noindent}
\newcommand{\ben}{\begin{enumerate}}
\newcommand{\een}{\end{enumerate}\par\noindent}
\newcommand{\beq}{\begin{equation}}
\newcommand{\eeq}{\end{equation}\par\noindent}
\newcommand{\beqa}{\begin{eqnarray*}}
\newcommand{\eeqa}{\end{eqnarray*}\par\noindent}
\newcommand{\beqn}{\begin{eqnarray}}
\newcommand{\eeqn}{\end{eqnarray}\par\noindent}

\def\jR{\begin{color}{DarkRed}}
\def\jB{\begin{color}{MidnightBlue}}
\def\jM{\begin{color}{magenta}}
\def\jC{\begin{color}{cyan}}
\def\jW{\begin{color}{white}}
\def\jBl{\begin{color}{black}}
\def\jG{\begin{color}{green}}
\def\jY{\begin{color}{yellow}}

\def\cR{\begin{color}{Crimson}}
\def\cB{\begin{color}{cyan}}
\def\cM{\begin{color}{magenta}}
\def\cC{\begin{color}{cyan}}
\def\cW{\begin{color}{white}}
\def\cBl{\begin{color}{black}}
\def\cG{\begin{color}{green}}
\def\cY{\begin{color}{yellow}}

\usepackage{enumitem}
\setlist[itemize]{noitemsep}
\setlist[enumerate]{noitemsep}

\begin{document}

\title{Any consistent coupling between classical gravity and quantum matter is fundamentally irreversible}

\author{Thomas D. Galley}
\email{thomas.galley@oeaw.ac.at}
\affiliation{Institute for Quantum Optics and Quantum Information, Austrian Academy of Sciences, Boltzmanngasse 3, 1090 Vienna, Austria}

\author{Flaminia Giacomini}
 \email{fgiacomini@phys.ethz.ch}
\affiliation{Institute for Theoretical Physics, ETH Z{\"u}rich, 8093 Z{\"u}rich, Switzerland
}

\author{John H. Selby}
 \email{john.h.selby@gmail.com}
\affiliation{ICTQT, University of Gda\'{n}sk, Wita Stwosza 63, 80-308 Gda\'{n}sk, Poland
}

\twocolumn[
\begin{@twocolumnfalse}
\begin{abstract}
When gravity is sourced by a quantum system, there is tension between its role as the mediator of a fundamental interaction, which is expected to acquire nonclassical features, and its role in determining the properties of spacetime, which is inherently classical. Fundamentally, this tension should result in breaking one of the fundamental principles of quantum theory or general relativity, but it is usually hard to assess which one without resorting to a specific model. Here, we answer this question in a theory-independent way using General Probabilistic Theories (GPTs). We consider the interactions of the gravitational field with a single matter  system, and derive a no-go theorem showing that when gravity is classical at least one of the following assumptions needs to be violated: (i) Matter degrees of freedom are described by fully non-classical degrees of freedom; (ii) Interactions between matter degrees of freedom and the gravitational field are reversible; (iii) Matter degrees of freedom back-react on the gravitational field. We argue that this implies that theories of classical gravity and quantum matter must be fundamentally irreversible, as is the case in the recent model of Oppenheim et al. Conversely if we require that the interaction between quantum matter and the gravitational field  {is} reversible, then the gravitational field must be non-classical. 
\end{abstract}
  \end{@twocolumnfalse}
]
\maketitle

\section{Introduction}

The gravitational field plays two roles. On the one hand, gravity is a fundamental interaction coupled to quantum matter, and as such it is natural to look for its quantum description. On the other hand, the gravitational field characterises the structure of spacetime, which is a classical concept coming from general relativity.  Reconciling these two roles when the source of gravity is quantum is one of the most profound conceptual challenges in finding a unified theory of gravity and matter.

Recently, the possibility of testing the gravitational field of a quantum source has attracted a lot of attention~\cite{bahrami2015gravity, anastopoulos2015probing, bose2017spin, marletto2017gravitationally, marletto2017we, carlesso2017cavendish, hall2018two, marletto2018can, belenchia2018quantum, belenchia2019information, christodoulou2019possibility, anastopoulos2020quantum, howl2020testing, marshman2020locality, Chevalier:2020uvv, krisnanda2020observable, marletto2020witnessing, galley_no_2020, pal2021experimental, Carney:2021yfw,universe8020058, danielson2021gravitationally, kent2021testing,  Christodoulou:2022vte, Huggett:2022uui, Christodoulou:2022knr, Danielson:2022tdw, Chen:2022wro, Martin-Martinez:2022uio, Overstreet:2022zgq}. No experiment using the technology currently available can test this scenario now~\cite{Aspelmeyer:2022fgc}, but these tests are expected to be 
feasible in the next  {couple of decades}, and 
would bring a major advancement to our understanding of gravity in regimes so far fully unexplored. On the theoretical side, the precise implications of these experiments are still debated~\cite{Christodoulou:2022vte}. From a fundamental point of view, however, this regime gives us the opportunity to test the fundamental principles of both quantum theory and general relativity together in a single scenario. Hence, it is crucial now to assess what the logical implications of combining the two theories in this regime are.

Here, we set to this task, and we ask whether we need to give up any of the fundamental principles of our theories. To achieve this goal, we use a theory-independent approach, analogous to  {that used in} Bell's theorem \cite{bell1964einstein}, 
via the framework of General Probabilistic Theories (GPTs) \cite{hardy2001quantum,barrett2007information} in contrast to earlier work which tends to consider specific, theory-dependent, models of interactions between matter and gravity,  {such as} the Schr\"odinger-Newton equation. In this  {theory independent} approach, we prove a no-go theorem, in the spirit of our recent work of Ref.~\cite{galley_no_2020}, that states that, if we want to keep a classical description of gravity and we allow quantum matter to source the gravitational field, then it is impossible to preserve the principle of reversibility of fundamental interactions. In general, our theorem shows under which conditions non-classical (including quantum) and classical degrees of freedom can couple in a consistent way. This is more general than previous work which was restricted to studying classical-quantum coupling~\cite{diosi_coupling_1998,caro_impediments_1999,diosi_quantum_2000,terno_inconsistency_2006,elze_linear_2012}.

We formally state the no-go theorem after introducing the relevant preliminaries of the theory independent framework which we work in. Informally however, the no-go theorem states that   {given a general classical system (in our case, the gravitational field)}, then a physical system coupled to it cannot at once (i) satisfy the superposition principle, (ii) interact reversibly with the  {classical system}, and (iii) back react on the  {classical system}.  {Formally, the theorem holds for finite-dimensional systems. However, operationally every physical system can be reduced to a finite dimensional system, because in the laboratory we can only perform a finite set of measurements.}

 {From our result, it is clear that if one wants to preserve these three principles, then the gravitational field ought to be nonclassical. Otherwise, if one} wants to hold to classical gravity, then  the only consistent ways to couple a quantum system with a classical system such that there is back-reaction from the quantum to the classical system are to either a) have an irreversible interaction as in Refs.~\cite{Oppenheim_post_2018, Oppenheim_gravitationally_2022, layton2022healthier} or b) have a fundamentally superselected  quantum system with only the classical ``which sector'' degrees of freedom interacting with the classical system. 

In the GPT literature similar theorems have been proved to study measurement theory~\cite{Heinosaari_no_2019}. In this case the classical system is a measuring device and the system being measured is non-classical. Since a required feature of a measurement interaction is that the state of the measuring device after the interaction is a function of the state of the system there must be back-reaction. Thus measurement of non-classical systems must be irreversible (i.e., involve disturbance in the language of~\cite{Heinosaari_no_2019}).

\section{General probabilistic theories: a theory independent framework}

A generalised probabilistic theory  is a mathematical object which precisely captures the operationally relevant content of a physical theory \cite{chiribella2010probabilistic},  {at least} for scenarios involving single agents and a fixed causal structure. That is, it captures only the information in the physical theory which is necessary and sufficient for making predictions about the outcomes of any such experiments that can be performed within the theory.  Operationally, the state of a system is characterised by finitely many measurements, and hence the state space of the system is finite dimensional. This entails that quantum systems have finite dimensional Hilbert spaces and  classical systems have finite configuration spaces. Alternatively one could accept the infinite dimensional nature of systems such as the gravitational field and instead understand the finite dimensional GPT description as a coarse grained version of the underlying infinite dimensional system.

A given physical theory may of course do more than  {describing the operational aspects}, it may also describe ontological or interpretational aspects. For example, a physical theory could be Bohmian mechanics \cite{bohm1952suggested} or Many-Worlds quantum theory \cite{everett2015theory}, but both would be associated to the same generalised probabilistic theory, namely, the standard formalism of quantum  {(}information {)} theory. It is therefore natural to assume that, whatever the underlying theory of nature actually is,  {there is a regime in which it is well described by a GPT\footnote{It is, however, often highly nontrivial to determine which GPT is associated to a given physical theory and there are often surprising results. For example, the GPT associated to quantum theory with a non-linear modification of the Schr\"odinger equation is actually a classical GPT ~\cite{mielnik_mobility_1980}, as the non-linearity allows for perfect distinguishability of all pure states. For some relevant models, such as the Reginatto-Hall model~\cite{reginatto2009quantum} the GPT description is not known, and as such our results do not immediately apply.}. This regime  { at least} includes all of the standard kinds of experiments which we perform, but, { the particular formalism that we use here} would exclude for example, experiments involving indefinite causal structure \cite{hardy2005probability,chiribella2009beyond,oreshkov2012quantum} (in which the pieces of apparatus locally have a well defined causal structure but in which no global causal structure exists between them) or extended Wigner's friend scenarios \cite{wigner1995remarks,frauchiger2018quantum,Bong_2020,e23080925,schmid2023review,yīng2023relating} (in which the experimenters themselves are treated as part of the experimental scenario). }  { The GPT framework that we present here can, however, also be extended to include these more general scenarios, for example~\cite{hardy2005probability,dariano2014determinism,selby2022time,wilson2022quantum,vilasini2019multi,ormrod2023theories}.} Any results that can be proven in this bare-bones operational framework therefore automatically apply to  {most physical theories} even if they are not themselves expressed in the language of GPTs.

More concretely, a GPT describes a set of preparations, transformations and measurements and encodes the probabilities of obtaining measurement outcomes given a circuit of preparations, transformations and measurements. These probabilities may be given by classical or quantum theory, or some different theory. 
In this section we do not provide a full introduction to GPTs but rather refer the reader to the works~\cite{barrett2005no,janotta2014generalized,plavala_general_2021}. Here, we introduce the notation and important definitions required for the main theorem.

In the following we denote by $\Omega$ the  {convex} set of states of a GPT system $\S$, $\E$ the set of effects and $\T$ the set of transformations. If $V$ is the  {finite dimensional} real vector space associated to $\S$ then $\Omega \subset V$\footnote{Where the symbol $\subset$ means ``is a convex subset of".}, $\E \subset V^*$ and any $t\in\T$ induces a linear transformation on $V$.
In a given GPT two systems $\S_1$ and $\S_2$ with associated vector spaces $V_1$ and $V_2$ compose to a third system $\S_3$ with associated vector space $V_3$ via a bilinear map $\otimes: V_1 , V_2 \to V_3$. When the theory is locally tomographic \cite{hardy2001quantum} (such as is the case for quantum and classical theory) $\otimes$ is the standard tensor product.

\begin{definition}[Reversible interaction]\label{def:R}
A reversible interaction between two GPT systems $\S_1$ and $\S_2$ with respective state spaces $\mS_1$ and $\mS_2$ is a linear map $\mI: \mS_1 \otimes \mS_2 \to \mS_1 \otimes \mS_2$ where there exists an inverse map   $\mI^{-1}: \mS_1 \otimes \mS_2 \to \mS_1 \otimes \mS_2$ such that $\mI^{-1} \circ \mI = \mathbb{I}_{\mS_1 \otimes \mS_2}$.
\end{definition}

\begin{definition}[Information flow]\label{def:IF}
Given an interaction $\mI$ between two GPT systems  $\S_1$ and $\S_2$ with input states $s_1$ and $s_2$ there is information flow from $\S_1$ to $\S_2$ if a change in state of system $\S_1$ before the interaction can lead to a change of the state of system $\S_2$ after the interaction.
\end{definition}

If we denote the choices of state preparation by the classical variables $s_1$ and $s_2$ and denote the outcomes of possible measurements after the interaction by the variables $e_1$ and $e_2$ this is equivalent to saying there is signalling from $1$ to $2$, that is, $p(e_2|s_1,s_2, \mI)$ depends non-trivially on $s_1$.

\begin{definition}[Finite dimensional classical system]\label{def:C}
	A classical system of dimension $n$ has a state space given by an $n$-vertex simplex $\Delta_n$ with associated vector space $V \cong \mathbb{R}^{n+1}$. Effects belong to the hypercube of $[0,1]$-valued linear functionals in $V^*\cong \mathbb{R}^{n+1}$ and transformations are given by stochastic linear maps. Classical systems compose with all other systems (both classical and non-classical) via the tensor product.
\end{definition}

 {Intuitively, this definition states that any two classical states can be perfectly distinguished with an operation belonging to the set of classical effects, and that any combination of two classical states leads at most to a classical mixture of probabilities (e.g.,\,no quantum interference is allowed). Finally, the standard tensor product composition rule ensures that no entangled states are generated when two classical systems are combined.}

\begin{definition}[Reducible GPT system]
	A GPT system $\S$, with associated vector space $V$, is reducible if and only if there is a decomposition  $V \cong \bigoplus_i V_i$  such that all states are convex combinations of states having support in a single block. That is, such that the state space $\Omega \cong \bigoplus_i \Omega_i$ where $\Omega_i$ is a convex set in $V_i$.
\end{definition}

A reducible GPT system is one that has a classical degree of freedom, and, hence, is at least partly classical. For example, the finite dimensional classical systems described above can be written as $\Delta_n = \underbrace{\Delta_1 \oplus \cdots \oplus \Delta_1}_{n}$ and, more generally, if we have a reducible system of the form $\underbrace{\Omega\oplus \cdots\oplus \Omega}_{n}$ then this is nothing but $\Omega\otimes \Delta_n$, that is, it factorises as a classical system, $\Delta_n$ and another system $\Omega$. Finally, any reducible system $\bigoplus_{i=1}^n \Omega_i$ can be thought of as a probabilistic mixture of the systems $\Omega_i$, where there is some label $i$ telling us which system we have. This  label cannot be prepared in `superpositions' and so is best thought of as a classical degree of freedom. In the case of quantum theory, reducible systems are those that have a non-trivial superselection rule, and the classical label tells us which superselection sector the state has been prepared in.  This motivates the following definition:

\begin{definition}[Fully nonclassical systems]\label{def:FNC} A fully nonclassical GPT system $\S$ is one that is irreducible.
\end{definition}

 Irreducible systems can be characterised by the fact that their identity transformations cannot be viewed as a coarse graining of other transformations, or, equivalently \cite{d2019information}, that there do not exist any non-disturbing measurements for the system. See also~\cite[Section 5.1]{Heinosaari_no_2019} for more discussion of reducible GPT systems.

\section{Results}

Our main technical result is the following no-go theorem, which follows directly from Theorem~\ref{Th:NonDisturbingMeasurement} proven in Appendix~\ref{App:Proof}.  {Moreover an algebraic version of the proof for the special case of quantum systems is given in Appendix~\ref{app:CQproof} for the reader who is less familiar with diagrammatic approaches.}

\begin{theorem}\label{cor:main}
Given two GPT systems $\S$ and $\G$ which interact via some interaction $\mI$ then at least one of the following conditions must be violated:

\begin{enumerate}[label=(\roman*)]
    \item The system $\S$ is fully non-classical (Def.~\ref{def:FNC})
    \item The interaction $\mI$ is reversible (Def.~\ref{def:R})
    \item There is information flow from system $\S$ to system $\G$ (Def.~\ref{def:IF})
    \item $\G$ is classical (Def.~\ref{def:C})
\end{enumerate}
\end{theorem}

Furthermore Theorem~\ref{Th:NonDisturbingMeasurement} shows that if $\S$ is reducible and $\G$  is classical, it is only the classical (i.e., the which-sector) information which can flow from $\S$ to a $\G$ via a reversible interaction $\mI$.

 \emph{Proof sketch:}  Suppose that conditions  $(i) - (iv)$ of the theorem hold and consider the following procedure. First evolve $\G$ and $\S$ by $\mI$, which by condition $(iii)$ constitutes a non-trivial measurement of $\S$ (see Example~\ref{ex:classical_meas}). Following this the state of $\G$ state can be copied onto another system $\G'$ (in the sense of Equation~\eqref{eq:classical_copy})  since $\G$ is classical by condition (iv). Subsequently the inverse $\mI^{-1}$ evolution (which exists by condition $(ii)$) is applied which returns  $\S$ and $\G$ to their initial states. The information gained about $\S$ during $\mI$ is not erased since it has been copied onto $\G'$. Hence this total procedure constitutes a disturbance free measurement of the irreducible system $\S$ (condition $(i)$) with non-trivial information gain. It follows from the information gain versus state disturbance trade-offs that this is impossible and therefore at least one of the assumptions $(i)-(iv)$ must be violated.

We can intuitively understand this theorem in the case of quantum theory by appealing to an example from Ref.~\cite{diosi_quantum_2000}. Consider a classical harmonic oscillator (system $\G$)  with free Hamiltonian $H_0 = \frac{1}{2}p^2$ interacting with a quantum spin system (system $\S$) via the interaction Hamiltonian $H_I = \kappa \hat \sigma_3 p$. The $x$ coordinate cannot satisfy  {$\dot{x} = \partial_p (H_0 + H_I) = p + \kappa \hat \sigma_3$, where the dot denotes the time derivative and $\partial_p$ stands for the partial derivative by $p$,} since  {$\hat \sigma_3$} is an  {operator}. Hence we might use the expectation value  {$\langle \hat \sigma_3 \rangle$} instead. This is then an interaction which is reversible and admits back-reaction, hence obeys conditions (ii) and (iii). However, this interaction 
\begin{quote}\emph{``implies that quantum expectations can be deduced with arbitrary precision from the measurement of the classical variables x and p."}~\cite{diosi_quantum_2000}.
\end{quote}
This means that either the classical system must inherit some of the quantum uncertainty (and no longer be classical) in order to prevent violation of the uncertainty principle of the quantum system or that the quantum system no longer satisfies the uncertainty principle, and so has inherited classical properties. In other words either condition (iv) is violated since $\G$ has become non-classical, or condition (i) is violated, since $\S$ has become classical.

Let us emphasise that the proof of the theorem only requires us to consider interactions between a classical system and a non-classical system -- in particular, we do not need any assumption on the properties of a nonclassical theory of gravity (e.g., Hilbert space factorisation, interactions, etc.). 

\subsection{ {Examples}}

 {We provide some simple examples of different types of interactions between systems and which of the assumptions of the theorem they meet. }

\begin{example}[Entangling description of a quantum measurement] \label{ex:ent_meas}
    { Consider two qubits $\S$ and $\G$ evolving according to a CNOT gate. This constitutes a $Z$ measurement of $\S$ by $\G$ when $\G$ is initialised in $\ket{0}$:
    \begin{align}
        (\alpha \ket{0}_\S + \beta \ket{1}_\S) \ket 0_\G \mapsto \alpha \ket{00}_{\S\G} + \beta \ket{11}_{\S\G}
    \end{align}
    Here, conditions (i), (ii) and (iii) are met while condition (iv) is not. To see that condition (iii) is met observe that the final state $|\alpha|^2 \ketbra{0}{0}_\G + |\beta|^2 \ketbra{1}{1}_\G$ of $\G$  depends on the initial state $\alpha \ket{0}_\S + \beta \ket{1}_\S$ of $\S$. }
\end{example}

\begin{example}[Measurement of a quantum system by a classical system]\label{ex:classical_meas}
    {We now consider a qubit $\S$ on which a classical two level system $\G$ (represented by a superselected qubit in the $Z$ basis) performs a $Z$ observable measurement. The two level system evolves according a decohered CNOT gate:
    \begin{align}
        \ketbra{\psi}{\psi} \otimes \ketbra{0}{0} \mapsto |\alpha|^2 \ketbra{00}{00}_{\S\G} + |\beta|^2 \ketbra{11}{11}_{\S\G} ,
    \end{align}
where $\ket{\psi} = \alpha \ket{0}_\S + \beta \ket{1}_\S$. Here, conditions  (i), (iii) and (iv) are met whilst condition (ii) is not.}
\end{example}

\begin{example}[Reversible measurement of super\-selection sector by classical systems]
 {Let us consider two spin $\frac{1}{2}$ systems with a total angular momentum super-selection rule; i.e. where there cannot be coherence between sectors with different total angular momentum. Using known decomposition rules for total angular momentum we have:
    \begin{align}
        \frac{1}{2} \otimes \frac{1}{2} \simeq 0 \oplus 1,
    \end{align}
    where the Hilbert space decomposes into $\C^1 \oplus \C^3$ corresponding to the anti-symmetric and symmetric subspaces of $\C^2 \otimes \C^2$. 
    A general density operator for this super-selected system is block diagonal $\rho = \rho_0 \oplus \rho_1$. We can define the following total angular momentum (i.e. ``which sector'') measurement by a classical two level system:}
    \begin{align}
         (\rho_0 \oplus \rho_1) \otimes  \ketbra{0}{0}
         \mapsto  (\rho_0  \otimes  \ketbra{0}{0})  \oplus (\rho_1  \otimes \ketbra{1}{1})
    \end{align}
\end{example} 

 {The above example is a specific instance of superselection sectors induced by the action of a group $G$ on a Hilbert space $\Hi$~\cite{RevModPhys.79.555}. Here, conditions (ii), (iii) and (iv) are met whilst condition (i) is not.}

\begin{example}[Semi-classical gravity (see Appendix~\ref{app:SN}]
    {In semi-classical gravity the stress energy tensor $T_{\mu \nu}$ in the Einstein equation is replaced by the expectation value of the stress energy tensor $\langle T_{\mu \nu} \rangle_{\ket \psi}$ for the matter field $\ket{\psi}$:
\begin{equation}\label{eq:Ricci_semi_classical_main}
R_{\mu\nu} + \frac{1}{2}g_{\mu\nu}R = \frac{8\pi G}{c^4} \bra{\psi} \hat{T}_{\mu\nu} \ket{\psi} .
\end{equation}
   From this it follows (under some additional assumptions) that $\ket{\psi}$ evolves according to the (non-linear) Schr{\" o}dinger-Newton equation~\cite{bahrami_schrodingernewton_2014}:
   \begin{equation} \label{eq:SchNew}
i \frac{d}{dt} \psi = \left[- \frac{\nabla^2}{2m}   -Gm^2 \int d^3r' \frac{|\psi(t, r')|^2}{|r-r'|}+ V \right]\psi  \, .
\end{equation}}
\end{example}

 {The key feature of semi-classical gravity (and semi-classical approximations more generally as described in Appendix~\ref{app:mean-field}) is that they induce dynamics described by a non-linear Schr{\" odinger} equation. Importantly `quantum' systems with such modified dynamics no longer have the same operational predictions as quantum theory, and as such do not correspond to the same GPT. It was shown in Ref.\,\cite{mielnik_mobility_1980} for the family of NLSE which are non-linear in $|\psi|^2$ that such systems are in fact fully classical, with states corresponding to probability measures over the projective Hilbert space of quantum states.  This is analogous to the `classicalisation' of the quantum spin system coupled reversibly to the classical harmonic oscillator discussed above. This then shows that semi-classical gravity, which claims to couple quantum matter with a classical gravitational field, violates condition (i) and therefore describes classical matter interacting with a classical gravitational field.}

 In semi-classical gravity the Sch{\"o}dinger-Newton equation appears  as a \emph{fundamental} description of quantum matter interacting with classical gravity~\cite{bahrami_schrodingernewton_2014}. 
 We note that non-linear Schr{\" o}dinger equations, including the Schr{\" o}dinger-Newton equation also appear in effective descriptions of quantum systems via mean-field approximations~\cite{bahrami_schrodingernewton_2014,breuer_theory_2002}. We discuss these in Appendix~\ref{app:mean-field} and note that whilst the theorem also applies to them, they are of less conceptual interest since they are understood to be useful approximations for describing interacting quantum systems rather than a fundamental description of quantum-classical interaction.

\section{ {Discussion}}

 {Let us discuss what a rejection of each of the conditions in Theorem~\ref{cor:main} entails:}

\vspace{0.2cm}

\noindent \textit{Condition (i).} Rejecting this condition would mean that any matter degrees of freedom which have a back-action on the gravitational field { (for example the position of a massive particle)} are modelled by reducible systems, for instance superselected quantum systems or classical systems. Since massive quantum particles are irreducible systems, this requires us to reject quantum theory as describing any matter degrees of freedom which back-react on the gravitational field. 
Essentially, rejecting condition (i) means rejecting quantum theory and the superposition principle.  {We note that there are two main ways in which a quantum system can `classicalise'. The first is via a superselection rule, so that superpositions of states in a given basis are not allowed. The second is that the projective Hilbert space of pure quantum states is in fact a classical configuration space and the mixed states are measures on this space. For a qubit which has pure states which are isomorphic to a sphere, the corresponding classical system consists of all measures on the sphere with pure states being Dirac measures~\footnote{ {This classical system appears in the Beltrametti-Bugajski model of quantum systems~\cite{beltrametti1995classical}.}}. In this case a pure state $\delta_{\frac{1}{\sqrt{2}}(\ket 0 + \ket 1)}$ is perfectly distinguishable from the states $\delta_{\ket 0}$ and $\delta_{\ket 1}$. }
This is the case, e.g.,~of the Schr\"odinger-Newton equation, discussed in Appendix~\ref{app:SN}. 

\vspace{0.2cm}

\noindent \textit{Condition (ii).} Rejecting this condition  {can lead} to the introduction of some sort of stochasticity, as in Refs.~\cite{Oppenheim_post_2018,Oppenheim_gravitationally_2022, carney2023strongly}.  For instance, in~\cite{Oppenheim_post_2018,Oppenheim_gravitationally_2022} a model of quantum matter coupled to classical gravity is proposed. The interaction leads to linear dynamics in the density matrix, unlike proposals such as the Schr\"odinger-Newton equation for instance, meaning that the matter degrees of freedom are fully described by quantum theory and therefore non-classical. The evolution is stochastic and is equivalent to an open quantum system description for the matter degrees of freedom. There is back-reaction of the matter degrees of freedom on the gravitational field. As such this proposal violates condition (ii), whilst meeting conditions (i) and (iii).

\vspace{0.2cm}

\noindent \textit{Condition (iii).} The case where there is no back-reaction from matter onto the gravitational field may emerge as a good approximation in certain situations, such as a low mass particle interacting gravitationally with a large static body.
However, while condition (iii) can be violated in certain approximations it would be in contradiction with a fundamental feature of gravity if the gravitational field was not influenced by matter.

\vspace{0.2cm}

\noindent  {\textit{Condition (iv).} In this case the gravitational field is non-classical. In the context of the current discussion on the low-energy nature of the gravitational field~\cite{bahrami2015gravity, anastopoulos2015probing, bose2017spin, marletto2017gravitationally, marletto2017we, carlesso2017cavendish, hall2018two, marletto2018can, belenchia2018quantum, belenchia2019information, christodoulou2019possibility, anastopoulos2020quantum, howl2020testing, marshman2020locality, Chevalier:2020uvv, krisnanda2020observable, marletto2020witnessing, galley_no_2020, pal2021experimental, Carney:2021yfw,universe8020058, danielson2021gravitationally, kent2021testing,  Christodoulou:2022vte, Huggett:2022uui, Christodoulou:2022knr, Danielson:2022tdw, Chen:2022wro, Martin-Martinez:2022uio, Overstreet:2022zgq}, the violation of this condition is the only one allowing one to preserve both reversibility and back-action when gravity interacts with quantum matter. The violation of this condition thus consists in a justification of the search for an (unknown) quantum gravity theory at higher energies, which might have very different features to those predicted in the low-energy regime. Whilst non-classicality of the gravitational field is consistent with the interaction being either reversible or irreversible we note that all the proposals the authors are aware of are reversible.}

\subsection{On reversibility as a feature of quantum matter interacting with the gravitational field}

 {Let us now apply the theorem to the question of whether the gravitational field needs to be quantized. We assume that matter is irreducibly quantum (condition (i)) and that it back-reacts on the gravitational field (condition iii).}

 {If one is committed to reversibility as a fundamental feature of the interaction between matter and the gravitational field, then one is forced to conclude that the gravitational field is non-classical. If one is not committed to reversibility then this leaves open the possibility of the gravitational field being classical.}

 {Theorem~\ref{cor:main} is not normative and gives no reason to prefer rejecting one assumption over another. Rather one must appeal to external arguments to justify different choices. There is ample discussion about how fundamental reversibility is in physics. On the one hand, quantum field theory (our best fundamental microscopic theory of matter) and general relativity (our best theory of the interaction between classical matter and space-time) are reversible. Hence, one may expect that a theory combining the two should preserve reversibility. On the other hand, we have potential instances of irreversibility in fundamental physical theories, such as measurements in quantum theory and thermodynamical processes. Even though one may argue that at a more fundamental level thermodynamics can be reduced to statistical mechanics, this reduction is controversial and the status of reversibility within statistical mechanics is still debated.}

Finally, we observe that reversibility is not straightforward to falsify, since the observation of an irreversible process does not preclude the possibility that it is a part of a larger reversible process.

\subsection{ Should the gravitational field be quantized?}

 {When studying reversible interactions between quantum matter and classical gravity Mielnik concluded that ``Either gravity is quantum or the electron is classical'' and therefore ``though it may be very difficult to quantize the gravitation, it is even more difficult not to do it''~\cite{mielnik1974}. Similarly~\cite{marletto2017we} states that ``everything must be quantized'' using the framework of constructor theory.  This framework assumes reversibility as well as the assumption that `every measurer of a Z observable' (which is the CNOT gate of Ex.\ref{ex:ent_meas} for quantum theory) can be used to perfectly discriminate states in the $X$ basis (which is met in quantum theory since the input states $\ket \pm \ket 0 $ are mapped to orthogonal states).  Moreover superselected systems are not considered since the non-classical system is assumed to have just two observables, which are incompatible }  {While both these results~\cite{marletto2017we,mielnik1974} form part of a larger debate on the necessity of quantizing the gravitational field, the present work makes precise exactly which assumptions are needed to construct such arguments, for instance by highlighting the key roles reversibility and superselection play. }

 {The results of this paper show that the above statements about the necessity of quantizing the gravitational field  are only true under the assumption of reversibility, and that moreover if one accepts super-selected quantum matter one can preserve both reversibility and classicality of the gravitational field. }

 {A natural extension of the results of this paper is to consider infinite dimensional systems. For an extension to infinite dimensional classical systems interacting with irreducible finite dimensional quantum systems the algebraic version of the proof in Appendix~\ref{app:CQproof}  can  informally be adapted by replacing $\sum_x \to \int_x dx$. The diagrammatic proof of Appendix~\ref{App:Proof} is expected to extend to the general case, however a fully rigorous extension  to infinite dimensional classical and non-classical systems is left to future work. }

\subsection{ {Classical gravity and quantum matter}}

If gravity is classical, at least one of the conditions (i)-(iii) has to fail. In summary, this implies the following.
\begin{center}
\centering
\begin{tabular}{|c|c|}
\hline 
Fail (i) & Reject  {irreducibly quantum matter}  \\ 
\hline 
Fail (ii) & Reject reversibility  \\ 
\hline 
Fail (iii) & Reject matter dependence of gravity  \\ 
\hline 
\end{tabular} 
\end{center}
Hence, a theory in which classical general relativity holds and matter is quantum necessarily needs to break the principle of reversibility. This means that the main feature of the model of \cite{Oppenheim_post_2018,Oppenheim_gravitationally_2022} which one could object to, namely stochasticity of the dynamics, is in fact inevitable for any model which seeks to combine quantum matter with classical general relativity.

\acknowledgments{The authors would like to thank Barbara \v{S}oda for helpful feedback on the manuscript  {and the referees for their comments which helped improve the manuscript}. T.D.G acknowledges support from the Austrian Science Fund (FWF), project P 33730-N. F.G. acknowledges support from the Swiss National Science Foundation via the Ambizione Grant PZ00P2-208885, from the ETH Zurich Quantum Center, and from the John Templeton Foundation, as part of the \href{https://www.templeton.org/grant/the-quantuminformation-structure-ofspacetime-qiss-second-phase}{‘The Quantum Information Structure of Spacetime, Second Phase (QISS 2)’ Project}. The opinions expressed in this publication are those of the authors and do not necessarily reflect the views of the John Templeton Foundation. T.D.G. and F.G. acknowledge support from Perimeter Institute for Theoretical Physics. Research at Perimeter Institute is supported in part by the Government of Canada through the Department of Innovation, Science and Economic Development and by the Province of Ontario through the Ministry of Colleges and Universities. J.H.S. was supported by the Foundation for Polish Science through IRAP project co-financed by EU within Smart Growth Operational Programme (contract no. 2018/MAB/5).}

\vspace{0.5cm}

\noindent \textbf{Author contributions:} All authors contributed equally to this work.

\bibliographystyle{quantum}
\bibliography{biblio}
\appendix

\onecolumngrid

\section{ Semi-classical approximations}\label{app:mean-field}

 {In general semi-classical approximations involve modelling some physical quantity by its expectation value. In some cases quantum fluctuations about the expectation value of the physical quantity are also included. Moreover these approximations are often accompanied by an assumption of locality, such as the Hartree-Fock approximation. }

\begin{example}[Mean field approximation]
   {  In a mean field approximation an observable $\hat O_i$ is replaced by its mean value and some fluctuation about this value $\hat O_i \approx \langle O_i \rangle + \delta\hat O_i$. A Hamiltonian of the form  $\hat H = \sum_i K_i + \sum_{\langle ij\rangle }\hat O_i \hat O_j$, where $\langle ij \rangle$ indicates nearest neighbours, in this approximation is $\hat H \approx  \sum_{\langle i j \rangle}\left(\langle \hat O_j \rangle \hat O_i + \langle \hat O_i \rangle \hat O_j - \langle \hat O_i \rangle \langle \hat O_j \rangle\right)$. The time evolution generated by this Hamiltonian is a non-linear Schrodinger equation, as can be seen by the presence of state dependent terms in the Hamiltonian.}
\end{example}

 {The term mean-field approximation is also used to describe the following approximation: first the Hartree-Fock approximation is made (multi-particle states are assumed to be product) and then one assumes that the Hamiltonian for a single system depends on the expectation value of some observable of the other particles (such as charge density). This leads to non-linear quantum master equations and non-linear Schr{\" o}dinger equations~\cite[Section 3.7]{breuer_theory_2002}, such as the Gross-Pitaevskii equation for describing bosons. }

 {The Hartree-Fock approximation and mean field limit can also be used in the case of quantum gravity. Assuming linearized quantum gravity (i.e., an effective quantum field-theoretic description of gravity in the weak-field limit) and making the assumptions described above one can derive a non-linear Sch{\"o}dinger equation (the Sch{\"o}dinger-Newton equation) as an \emph{effective} description of the zeroth-order dynamics of N particles interacting with a quantum gravitational field in the aforementioned regime~\cite[Section 2.1]{bahrami_schrodingernewton_2014}. }

\section{Schr\"odinger-Newton} \label{app:SN}

The Schr{\" o}dinger-Newton equation~\cite{bahrami_schrodingernewton_2014}  appears in semi-classical gravity, where spacetime is described using the classical general relativistic framework and only matter degrees of freedom are quantized. Einstein's equations describe the coupling between spacetime and matter in general relativity, and can be generalised to quantum matter by letting the expectation value of the stress energy tensor couple to the gravitational field:
\begin{equation}\label{eq:Ricci_semi_classical}
R_{\mu\nu} + \frac{1}{2}g_{\mu\nu}R = \frac{8\pi G}{c^4} \bra{\psi} \hat{T}_{\mu\nu} \ket{\psi} ,
\end{equation}
From this semi-classical Einstein equation one can derive the Schr\"odinger-Newton equation describing the evolution of the matter degrees of freedom~\cite{bahrami_schrodingernewton_2014}:
\begin{equation} \label{eq:SchNew}
i \frac{d}{dt} \psi = \left[- \frac{\nabla^2}{2m}   -Gm^2 \int d^3r' \frac{|\psi(t, r')|^2}{|r-r'|}+ V \right]\psi  \, .
\end{equation}
This setup seems to at first contradict our theorem. Namely matter is described by an irreducible quantum system, the interaction is reversible, there is back-reaction from the matter degrees of freedom onto the gravitational field and the gravitational field is classical.

This apparent contradiction is resolved once one observes that the non-linear dynamics described by the Schr\"odinger Newton equation allows one to carry out new non-quantum measurements on the matter system. These new measurements allow one to distinguish arbitrary pure states, entailing that the system is essentially classical. Therefore the Schr\"odinger-Newton model violates condition 1. of the theorem since it describes matter degrees of freedom as classical~\cite{mielnik_mobility_1980}.

\section{Mielnik's generalized quantum mechanics}
\label{app:Mielnik}

In his 1974 paper ``Generalized quantum mechanics''~\cite{mielnik1974} Mielnik studied modifications to the Schr\"odinger equation induced by `subtle' interactions between a quantum system and a classical gravitational field. In the language of our theorem this corresponds to reversible interactions with back-reaction. In the light of our no-go theorem, Mielnik showed that imposing Conditions ii), iii) and that the gravitational field is classical led to a change in the nature of the quantum system. The new interactions allowed one to measure non-quadratic observables (i.e. non-quantum observables) which lead to the system being effectively classical, as in the case of the Schr\"odinger-Newton equation above. This led him to the following hypothesis: ``Either gravity is quantum or the electron is classical''. 

Our theorem allows us to turn his hypothesis into a more precise statement: ``Under the assumption of reversibility, either gravity is non-classical or the electron (i.e., matter) is a fundamentally super-selected (including classical) system''.

More specifically, if we assume that the position of the electron back-reacts on the gravitational field (as is predicted by Einstein's field equation) and that the interaction is reversible, then either gravity is non-classical or the position of the electron cannot be put into superpositions. A similar story can be told for the energy of the electron, or indeed, any other degree of freedom that we expect to back-react on the gravitational field. Note that this does not rule out the existence of non-classical degrees of freedom so long as they do not backreact on the gravitational field.

There exist other models of reversible interactions with back reaction between a classical system and a quantum system such as~\cite{peres_hybrid_2001} where the quantum degree of freedom inherits classical features. Theorem~\ref{Th:NonDisturbingMeasurement} shows that this is inevitable (unless one allows that classical system to become non-classical).

\section{Constraining interactions between classical and non-classical systems}\label{App:Proof}


We now give the theorem constraining the possible interactions between  
non-classical systems and classical systems which we used to obtain Theorem~\ref{cor:main} in the main text.

\begin{theorem}\label{Th:NonDisturbingMeasurement}
	There is no reversible interaction between a classical system, $\G$, and 
fully non-classical GPT system, $\S$ with information flow from the GPT system to the classical system. That is, any such interaction $\mathtt{R}$ must satisfy the no-signalling condition
\beq
\InputIfFileExists{Diagrams/Rcalc5.tikz}{}{\input{./figures/Diagrams/Rcalc5.tikz}}
\quad = \quad
\InputIfFileExists{Diagrams/rCalc7.tikz}{}{\input{./figures/Diagrams/rCalc7.tikz}}.
\eeq
Moreover, in the case that the  system $\S$ is reducible, then it is only the label of the sector which can influence the classical system in a reversible interaction. That is, it satisfies
\beq
\InputIfFileExists{Diagrams/Rcalc5.tikz}{}{\input{./figures/Diagrams/Rcalc5.tikz}}
\quad = \quad
\InputIfFileExists{Diagrams/rCalc8.tikz}{}{\input{./figures/Diagrams/rCalc8.tikz}},
\eeq
where $M$ is the non-disturbing measurement for $\S$ which measures the which-sector information.
\end{theorem}

In the case of quantum theory this theorem shows that the only interactions between classical and non-superselected quantum systems where the quantum system influences the classical system are irreversible. Moreover, if we do have superselected quantum systems, where information other than the which-sector information to influences the classical system, then this too must be governed by an irreversible interaction.

 {Note that we expect the same proof to hold for the case of infinite dimensional classical systems $\mathsf{G}$. On a formal level, what we need is to identify the relevant symmetric monoidal category for describing a generalised probabilistic theory coupled to infinite dimensional classical systems, at which point exactly the same proof that we give below should hold.}

\subsection{Proof of Theorem~\ref{Th:NonDisturbingMeasurement} for irreducible systems} \label{sec:mainproof}
\proof
Let us assume that we have a classical system $\G$, which in our key application will correspond to a classical gravitational field, and a fully non-classical system $\S$, which in our key application will correspond to the matter degree of freedom.
Then, let us assume that these two systems are interacting via a reversible interaction $\R$ which we diagrammatically denote as:
\beq
\InputIfFileExists{Diagrams/interaction.tikz}{}{\input{./figures/Diagrams/interaction.tikz}},
\eeq
where reversibility means that
there exists a physical transformation $\R^{-1}$ such that:
\beq
\InputIfFileExists{Diagrams/rev1.tikz}{}{\input{./figures/Diagrams/rev1.tikz}}
=
\InputIfFileExists{Diagrams/rev2.tikz}{}{\input{./figures/Diagrams/rev2.tikz}}
=
\InputIfFileExists{Diagrams/rev3.tikz}{}{\input{./figures/Diagrams/rev3.tikz}}
\eeq
At least in principle, there exist classical measurements, $\m$, of $\G$ which are non-disturbing\footnote{This does not necessarily need to be a `direct' measurement of the field, but could constitute a complicated `indirect' procedure by which some test particle interacts with the field and then some property of the test particle is measured to give the result.}.
We denote these as:
\beq\label{eq:leak}
\InputIfFileExists{Diagrams/ndm.tikz}{}{\input{./figures/Diagrams/ndm.tikz}}
\eeq
where $\sC$ is some classical `pointer' degree of freedom which records the measurement outcome.  In the case that $\G$ is a finite dimensional classical system, then there is a single non-disturbing measurement which perfectly records the state of $\G$ into $\sC$, we, however, do not need to make this assumption here.
The fact that this is a non-disturbing measurement is encoded in the fact that if we discard $C$ the resulting transformation on $F_g$ is simply the identity. {Such a transformation is known as a \emph{leak} in the process-theory literature \cite{selby2017leaks,selby2018reconstructing,coecke2017two}, in particular, one where the leaked system, $C$, is classical. Diagrammatically, these are characterised by the following equation:
\beq
\InputIfFileExists{Diagrams/ndm1.tikz}{}{\input{./figures/Diagrams/ndm1.tikz}}\label{eq:leaktrace}
= \quad %
\begin{tikzpicture}
	\begin{pgfonlayer}{nodelayer}
		\node [style=none] (0) at (0, -0.75) {};
		\node [style=right label] (1) at (0, -0.5) {$\G$};
		\node [style=none] (2) at (0, 2.25) {};
		\node [style=none] (3) at (0, 2.25) {};
	\end{pgfonlayer}
	\begin{pgfonlayer}{edgelayer}
		\draw (3.center) to (0.center);
	\end{pgfonlayer}
\end{tikzpicture}
}
\eeq
where
$\begin{tikzpicture}[scale=.5]
	\begin{pgfonlayer}{nodelayer}
		\node [style=upground] (0) at (0, .5) {};
		\node [style=none] (1) at (0, 0.25) {};
		\node [style=none] (2) at (0, -0.25) {};
	\end{pgfonlayer}
	\begin{pgfonlayer}{edgelayer}
		\draw (1.center) to (2.center);
	\end{pgfonlayer}
\end{tikzpicture}
$
represents throwing away/discarding/ignoring the physical system, $\sC$.  In quantum theory this is represented by the trace and in classical probability theory by marginalisation. In particular,
\beq
\begin{tikzpicture}
	\begin{pgfonlayer}{nodelayer}
		\node [style=upground] (0) at (0, 0.75) {};
		\node [style=none] (1) at (0, 0.5) {};
		\node [style=point] (2) at (0, -0.5) {$\P$};
		\node [style=right label] (3) at (0, 0) {$\sC$};
	\end{pgfonlayer}
	\begin{pgfonlayer}{edgelayer}
		\draw (1.center) to (2);
	\end{pgfonlayer}
\end{tikzpicture}
}
\eeq
therefore represents the normalisation of the state $\P$ and hence any physical state must satisfy:
\beq
\begin{tikzpicture}
	\begin{pgfonlayer}{nodelayer}
		\node [style=upground] (0) at (0, 0.75) {};
		\node [style=none] (1) at (0, 0.5) {};
		\node [style=point] (2) at (0, -0.5) {$\P$};
		\node [style=right label] (3) at (0, 0) {$\sC$};
	\end{pgfonlayer}
	\begin{pgfonlayer}{edgelayer}
		\draw (1.center) to (2);
	\end{pgfonlayer}
\end{tikzpicture}
}
=
1
\eeq
Similarly, any physical transformation must preserve this normalisation, hence we have that:
\beq
\begin{tikzpicture}
	\begin{pgfonlayer}{nodelayer}
		\node [style=upground] (0) at (0.5, 1.25) {};
		\node [style=none] (1) at (0.5, 1) {};
		\node [style=small box] (2) at (0.5, 0) {$\f$};
		\node [style=none] (3) at (0.5, -1) {};
	\end{pgfonlayer}
	\begin{pgfonlayer}{edgelayer}
		\draw (1.center) to (2);
		\draw (2) to (3.center);
	\end{pgfonlayer}
\end{tikzpicture}
}
=
\begin{tikzpicture}
	\begin{pgfonlayer}{nodelayer}
		\node [style=upground] (0) at (0.5, 1) {};
		\node [style=none] (1) at (0.5, 0.75) {};
		\node [style=none] (2) at (0.5, -0.25) {};
	\end{pgfonlayer}
	\begin{pgfonlayer}{edgelayer}
		\draw (1.center) to (2.center);
	\end{pgfonlayer}
\end{tikzpicture}
}.
\eeq

Now, consider the following diagram which we can construct out of these elements:
\beq\label{eq:composite1}
\InputIfFileExists{Diagrams/constNDM.tikz}{}{\input{./figures/Diagrams/constNDM.tikz}}.
\eeq
For some initial state of the field, $\s$, and ignoring the final state of the field, we have the following:
\beq \label{eq:composite2}
\InputIfFileExists{Diagrams/constNDM2.tikz}{}{\input{./figures/Diagrams/constNDM2.tikz}}.
\eeq
It is then simple to see that the above diagram describes a procedure that implements a non-disturbing measurement on the system $\S$, i.e., it defines a leak for the system, as:
\beq\label{eq:nondist}
\InputIfFileExists{Diagrams/constNDM3.tikz}{}{\input{./figures/Diagrams/constNDM3.tikz}}\quad = \quad %
\InputIfFileExists{Diagrams/constNDM4.tikz}{}{\input{./figures/Diagrams/constNDM4.tikz}}\quad = \quad %
\InputIfFileExists{Diagrams/constNDM5.tikz}{}{\input{./figures/Diagrams/constNDM5.tikz}}\quad = \quad %
\InputIfFileExists{Diagrams/constNDM6.tikz}{}{\input{./figures/Diagrams/constNDM6.tikz}}\quad = \quad %
\InputIfFileExists{Diagrams/constNDM7.tikz}{}{\input{./figures/Diagrams/constNDM7.tikz}}\quad = \quad %
\InputIfFileExists{Diagrams/constNDM8.tikz}{}{\input{./figures/Diagrams/constNDM8.tikz}}\quad = \quad %
\InputIfFileExists{Diagrams/constNDM9.tikz}{}{\input{./figures/Diagrams/constNDM9.tikz}}
\eeq
where the second equality uses the fact that $\m$ is non-disturbing for the field, the fourth reversibility of $\R$, and the sixth is the normalisation of the field state $\s$.

Now, it is a well known result that such non-disturbing measurements for irreducible GPT systems are trivial \cite{d2019information}, that is, that they factorise as the identity on the irreducible system and some fixed preparation of the classical pointer variable, that is:
\beq\label{eq:proof1}
\InputIfFileExists{Diagrams/constNDM2.tikz}{}{\input{./figures/Diagrams/constNDM2.tikz}}
\quad
=
\quad
\InputIfFileExists{Diagrams/trivialNDM.tikz}{}{\input{./figures/Diagrams/trivialNDM.tikz}}
\eeq
where this state $\mathtt{\chi_s^m}$ can depend on the initial state of the field $\s$ and the non-disturbing measurement of the field $\m$. We can compute this state to be given by:
\beq\label{eq:pointer1}
\begin{tikzpicture}
	\begin{pgfonlayer}{nodelayer}
		\node [style=point] (0) at (0, -0.75) {$\mathtt{\chi_m^s}$};
		\node [style=none] (1) at (0, 2) {};
		\node [style=right label] (2) at (0, 1.75) {$\sC$};
	\end{pgfonlayer}
	\begin{pgfonlayer}{edgelayer}
		\draw (1.center) to (0);
	\end{pgfonlayer}
\end{tikzpicture}
}
\quad
=
\quad
\InputIfFileExists{Diagrams/chi2.tikz}{}{\input{./figures/Diagrams/chi2.tikz}}
\quad
=
\quad
\InputIfFileExists{Diagrams/chi3.tikz}{}{\input{./figures/Diagrams/chi3.tikz}}
\quad
=
\quad
\InputIfFileExists{Diagrams/chi4.tikz}{}{\input{./figures/Diagrams/chi4.tikz}},
\eeq
which is independent of the arbitrarily chosen initial state $\mathtt{\rho}$. The first equality follows from normalisation of $\mathtt{\rho}$, the second from Eq.~\eqref{eq:proof1}, and the final one from the fact that $\R^{-1}$ is a physical transformation and so preserves normalisation.

Hence, we can rewrite eq.~\eqref{eq:proof1} as:
\beq\label{eq:new1}
\InputIfFileExists{Diagrams/constNDM2.tikz}{}{\input{./figures/Diagrams/constNDM2.tikz}}
\quad
=
\quad
\InputIfFileExists{Diagrams/constNDM9.tikz}{}{\input{./figures/Diagrams/constNDM9.tikz}}
\InputIfFileExists{Diagrams/chi4.tikz}{}{\input{./figures/Diagrams/chi4.tikz}}.
\eeq

Now, if we trace out the irreducible system, $\S$ in Eq.~\eqref{eq:new1}, we obtain
\beq%
\InputIfFileExists{Diagrams/Rcalc1.tikz}{}{\input{./figures/Diagrams/Rcalc1.tikz}}
\quad = \quad%
\InputIfFileExists{Diagrams/discS.tikz}{}{\input{./figures/Diagrams/discS.tikz}}%
\InputIfFileExists{Diagrams/chi4.tikz}{}{\input{./figures/Diagrams/chi4.tikz}} 
\eeq
which, using the fact that $\R^{-1}$ is physical and hence preserves normalisation gives
\beq%
\InputIfFileExists{Diagrams/Rcalc2.tikz}{}{\input{./figures/Diagrams/Rcalc2.tikz}}
\quad = \quad%
\InputIfFileExists{Diagrams/discS3.tikz}{}{\input{./figures/Diagrams/discS3.tikz}}%
\InputIfFileExists{Diagrams/chi41.tikz}{}{\input{./figures/Diagrams/chi41.tikz}} 
\eeq
Now, this is true for all states $\s$ of the field, hence we have
\beq
\InputIfFileExists{Diagrams/Rcalc3.tikz}{}{\input{./figures/Diagrams/Rcalc3.tikz}}\quad = \quad %
\InputIfFileExists{Diagrams/discS3.tikz}{}{\input{./figures/Diagrams/discS3.tikz}} %
\InputIfFileExists{Diagrams/rCalc4.tikz}{}{\input{./figures/Diagrams/rCalc4.tikz}}.
\eeq
Similarly, this holds for all non-disturbing measurements of the field $\m$, and as these suffice for tomography of the classical system $\G$ hence we can rewrite this as:
\begin{align}
\InputIfFileExists{Diagrams/Rcalc5.tikz}{}{\input{./figures/Diagrams/Rcalc5.tikz}}
\quad &= \quad
\InputIfFileExists{Diagrams/discS3.tikz}{}{\input{./figures/Diagrams/discS3.tikz}}%
\InputIfFileExists{Diagrams/rCalc6.tikz}{}{\input{./figures/Diagrams/rCalc6.tikz}}\\
\quad &=: \quad
\InputIfFileExists{Diagrams/rCalc7.tikz}{}{\input{./figures/Diagrams/rCalc7.tikz}}.
\end{align}
This is the definition of what it means for an interaction $\R$ to be no-signalling from $\S$ to $\G$. In other words, the state of the system $\S$ before the interaction has no influence on the state of the classical gravitational field $\G$ after the interaction.
\endproof

\subsection{Proof of Theorem \ref{Th:NonDisturbingMeasurement} for reducible systems}\label{proof:generalTheorem}

Up to Eq.~\ref{eq:proof1} the proof is identical, as that is the first point where we have invoked irreducibility of $\S$, the rest of the proof then proceeds as follows: 
\proof
Now, it is a well known result that such non-disturbing measurements for  GPT systems reveal only which-sector information \cite{d2019information}, that is, that they can be written as:
\beq\label{eq:proofnew1}
\InputIfFileExists{Diagrams/constNDM2.tikz}{}{\input{./figures/Diagrams/constNDM2.tikz}}
\quad
=
\quad
\InputIfFileExists{Diagrams/nontrivialNDM.tikz}{}{\input{./figures/Diagrams/nontrivialNDM.tikz}}
\eeq
where $\mathtt{M}$ is the non-disturbing measurement that reveals the which-sector information, that is, it satisfies:
\beq
\InputIfFileExists{Diagrams/MDef.tikz}{}{\input{./figures/Diagrams/MDef.tikz}}\quad = \quad %
\InputIfFileExists{Diagrams/MDef2.tikz}{}{\input{./figures/Diagrams/MDef2.tikz}}
\eeq
for all $i$ and $\rho_i \in \Omega_i$.
The process $\mathtt{\Sigma_s^m}$ is a stochastic map (which can depend on the initial state of the field $\s$ and the non-disturbing measurement of the field $\m$) processing the which-sector information. We can compute this stochastic map by looking at its action on point distributions as:
\beq
\InputIfFileExists{Diagrams/Sigma1.tikz}{}{\input{./figures/Diagrams/Sigma1.tikz}}
\quad
=
\quad
\InputIfFileExists{Diagrams/Sigma2.tikz}{}{\input{./figures/Diagrams/Sigma2.tikz}}
\quad
=
\quad
\InputIfFileExists{Diagrams/Sigma3.tikz}{}{\input{./figures/Diagrams/Sigma3.tikz}}
\quad
=
\quad
\InputIfFileExists{Diagrams/Sigma4.tikz}{}{\input{./figures/Diagrams/Sigma4.tikz}},
\eeq
where $\mathtt{\rho_i}$ is an arbitrarily chosen state in sector $\Omega_i$. The first equality follows from the definition of $\mathtt{M}$ together with the normalisation of $\mathtt{\rho_i}$, the second from Eq.~\eqref{eq:proofnew1}, and the final one from the fact that $\R^{-1}$ is a physical transformation and so preserves normalisation.

Hence, we can rewrite eq.~\eqref{eq:proofnew1} as:
\beq\label{eq:newnew1}
\InputIfFileExists{Diagrams/constNDM2.tikz}{}{\input{./figures/Diagrams/constNDM2.tikz}}
\quad
=
\quad \sum_i %
\InputIfFileExists{Diagrams/general1.tikz}{}{\input{./figures/Diagrams/general1.tikz}}
\quad
=
\quad \sum_i
\InputIfFileExists{Diagrams/general2new.tikz}{}{\input{./figures/Diagrams/general2new.tikz}}
\InputIfFileExists{Diagrams/Sigma4new.tikz}{}{\input{./figures/Diagrams/Sigma4new.tikz}},
\eeq
where in the first step we have taken Eq.~\eqref{eq:proofnew1} and introduced a resolution of the identity on the RHS, and in the second we have used the above characterisation of $\mathtt{\Sigma_s^m}$.

Now, if we trace out the system, $\S$ in Eq.~\eqref{eq:newnew1}, we obtain
\beq%
\InputIfFileExists{Diagrams/Rcalc1.tikz}{}{\input{./figures/Diagrams/Rcalc1.tikz}}
\quad = \quad\sum_i %
\InputIfFileExists{Diagrams/general3new.tikz}{}{\input{./figures/Diagrams/general3new.tikz}}
\InputIfFileExists{Diagrams/Sigma4new.tikz}{}{\input{./figures/Diagrams/Sigma4new.tikz}}
\eeq
which, using the fact that $\R^{-1}$ is physical and hence preserves normalisation gives
\beq%
\InputIfFileExists{Diagrams/Rcalc2.tikz}{}{\input{./figures/Diagrams/Rcalc2.tikz}}
\quad = \quad\sum_i %
\InputIfFileExists{Diagrams/general3new.tikz}{}{\input{./figures/Diagrams/general3new.tikz}}
\InputIfFileExists{Diagrams/Sigma4new.tikz}{}{\input{./figures/Diagrams/Sigma4new.tikz}}
\eeq
Now, this is true for all states $\s$ and measurements $\m$ of the field, hence we have:
\begin{align}
\InputIfFileExists{Diagrams/Rcalc5.tikz}{}{\input{./figures/Diagrams/Rcalc5.tikz}}
\quad &= \quad\sum_i %
\InputIfFileExists{Diagrams/general3.tikz}{}{\input{./figures/Diagrams/general3.tikz}}
\InputIfFileExists{Diagrams/general4.tikz}{}{\input{./figures/Diagrams/general4.tikz}}\\
\quad &=: \quad
\InputIfFileExists{Diagrams/rCalc8.tikz}{}{\input{./figures/Diagrams/rCalc8.tikz}}.
\end{align}
This means it is only the variation in the which-sector information of $S$ which signals to $\G$.
\endproof

\section{Algebraic version of the proof for classical-quantum systems}

\subsection{Algebraic proof for irreducible systems}\label{app:CQproof}

 {For the reader who is not familiar with diagrammatic notation and GPTs we offer an algebraic version of the proof of Theorem~\ref{Th:NonDisturbingMeasurement} for irreducible systems in the case where the non-classical system is quantum. }

 This proof broadly follows the diagrammatic proof of Section~\ref{sec:mainproof} and we refer to the diagrammatic version of equations so that the interested reader can translate between the diagrammatic approach and the algebraic approach.

 Let $\S$ be a non-super-selected quantum system $\C^d$ and $\G$ a finite $n$-dimensional classical system. We represent $\G$ using diagonal density operators on $\C^n$.

 A general classical quantum state is of the following form:
\begin{align}
    \tilde \rho_{\G\S} = \sum_x p(x) \ketbra{x}{x}_\G \otimes \rho(x)_\S ,
\end{align}
where $p(x)$ is some probability distribution and $\rho(x)_\S$ a density operator.
The interaction $R:\S \G \to \S \G $ is reversible hence there exists a transformation $R^{-1}:\S \G \to \S \G $ such that 
\begin{align}
    R^{-1}(R^{-1}(\cdot)) = \I_{\G\S} ,
\end{align}
with $\I_{\G\S}$ the identity super-operator on the classical quantum states. 

 {$R$ is assumed to be signalling from $\S \to \G$ hence for some input states $\rho_\S$ and $\sum_x p(x) \ketbra{x}{x}_\G$ the output state is:
\begin{align}
    R \left(\sum_x p(x) \ketbra{x}{x}_\G \otimes \rho_\S\right) = \sum_x q(x) \ketbra{x}{x} \otimes \sigma_\S(x) ,
\end{align}
where the reduced state $\sum_x q(x) \ketbra{x}{x}_\sC$ depends on the input state $\rho_\S$.}

The leak of Equation~\eqref{eq:leak} in the case where the measurement is the canonical measurement, i.e. the measurement in the $\ketbra{x}{x}_\G$ basis is given by the following map:
\begin{align}
    m&: \G \to \sC\G \\
    m&:: \sum_x p(x) \ketbra{x}{x}_\G \mapsto \sum_x p(x) \ketbra{x}{x}_\G  \otimes \ketbra{x}{x}_\sC  , \label{eq:classical_copy}
\end{align}
where $\sC$ is a classical system of the same dimension as $\G$. This operation can be understood as a sort of classical copy operation since the reduced state on $\sC$ after the transformation $m$ is equal to the initial state of $\G$. We stress that the diagrammatic proof is more general and applies to any leak, not the one corresponding to the canonical measurement only.

We observe that tracing out $\sC$ after $m$ is just the identity on $\G$:
\begin{align}
    \Tr_\sC\left(m\left(\sum_x p(x) \ketbra{x}{x}_\G\right)\right) = \Tr_\sC\left(\sum_x p(x) \ketbra{x}{x}_\G  \otimes \ketbra{x}{x}_\sC \right) = \sum_x p(x) \ketbra{x}{x}_\G ,
\end{align}
showing that this obeys the condition of Equation~\eqref{eq:leaktrace}.
The diagram of Equation~\eqref{eq:composite1} is translated to:

\begin{align}
    E&: \S\G \to \S\G\sC\\
    E&::\rho_{\G\S} \mapsto (R^{-1} \otimes \I_\sC) \left(m(R(\rho_{\G\S}))\right) . 
\end{align}
The interaction of Equation~\eqref{eq:composite2} is obtained from $E$ by fixing an initial state $s_\G = \sum_x p(x) \ketbra{x}{x}_\G$ of $\G$ and tracing  out the  state of $\G$ after $E$, this defines a new map:
\begin{align}
    E_{s_\G}&: \S \to \S\sC \\
    E_{s_\G}&::  \rho_\S \mapsto \Tr_\G\left(E\left(\sum_x p(x) \ketbra{x}{x} \otimes \rho_\S\right)\right)  .
\end{align}
We see that this map is a non-demolition measurement of $\S$ since it takes $\S$ to some joint state on $\S\sC$ where $\sC$ is the classical pointer system. Following Equation~\eqref{eq:nondist} we can show that this measurement is non-disturbing:
\begin{align}
    \Tr_\sC(E_{s_\G}(\cdot))&: \S \to \S \\
    \Tr_\sC(E_{s_\G}(\cdot))&:: \rho_\S \mapsto \Tr_{\G\sC}\left(E\left(\sum_x p(x) \ketbra{x}{x} \otimes \rho_\S\right)\right)\\
    =& \Tr_{\G\sC}\left((R^{-1} \otimes \I_\sC) \left(m \left(U_{\G \S} \left(\sum_x p(x) \ketbra{x}{x} \otimes \rho_\S \right) U^\dagger_{\G \S}\right)\right)\right) \\
    =& \Tr_{\G\sC}\left((R^{-1} \otimes \I_\sC) \left(m  \left(\sum_x q(x)\ketbra{x}{x} \otimes \sigma_\S(x) \right)\right) \right)\\
    =& \Tr_{\G\sC}\left((R^{-1} \otimes \I_\sC)\left(\sum_x q(x)\ketbra{x}{x} \otimes \sigma_\S(x) \otimes \ketbra{x}{x}_\sC\right)\right)\\
    =& \Tr_\G \left(R^{-1} \left(\Tr_\sC\left(\sum_x q(x)\ketbra{x}{x} \otimes \sigma_\S(x) \otimes \ketbra{x}{x}_\sC\right)\right)\right) \\
    =& \Tr_\G \left(R^{-1} \left(\sum_x q(x)\ketbra{x}{x} \otimes \sigma_\S(x) \right)\right) \\
    =& \Tr_\G \left(\sum_x p(x) \ketbra{x}{x}_\G \otimes \rho_\S\right) \\
    =& \rho_\S , 
\end{align}
where we have used the fact that $\Tr_{\G \sC}\left((R^{-1} \otimes \I_\sC)(\cdot)\right) = \Tr_\G \left((R^{-1} (\Tr_\sC(\cdot)\right) $. This is the first step in Equation~\eqref{eq:nondist}.

Let us compute the state of the pointer system $\sC$ after the interaction by tracing out the state of $\G\S$:
\begin{align}
\Tr_{\G\S}(E_{s_\G}(\cdot)) & : \S \to \sC \\
        \Tr_{\G\S}(E_{s_\G}(\cdot)) & ::  \rho_\S \mapsto \Tr_{\G\S}\left(E\left(\sum_x p(x) \ketbra{x}{x} \otimes \rho_\S\right)\right)\\
    =& \Tr_{\G\S}\left((R^{-1} \otimes \I_\sC) \left(m \left(U_{\G \S} \left(\sum_x p(x) \ketbra{x}{x} \otimes \rho_\S \right) U^\dagger_{\G \S}\right)\right)\right)  ,
\end{align}
and we use the fact that $R^{-1}$ preserves normalisation, so that $\Tr_{\S\G}(R^{-1} (\rho_{\S\G})) = \Tr_{\S\G}(\rho_{\S\G}) $ (the final step of Equation~\eqref{eq:pointer1}), then
\begin{align}
\Tr_{\G\S}(E_{s_\G}(\rho_\S))
    &= \Tr_{\G\S} \left(m \left(U_{\G \S} \left(\sum_x p(x) \ketbra{x}{x} \otimes \rho_\S \right) U^\dagger_{\G \S}\right)\right) \\
     &= \Tr_{\G\S}\left(m \left(\sum_x q(x)\ketbra{x}{x} \otimes \sigma_\S(x)\right)\right)\\
    &= \Tr_{\G\S}\left(\sum_x q(x)\ketbra{x}{x} \otimes \sigma_\S(x) \otimes \ketbra{x}{x}_\sC\right)\\
    &= \sum_x q(x)  \ketbra{x}{x}_\sC  . 
\end{align}
Observe that since we assume that there is information flow from $\S \to \G$ during the interaction $R$, this implies that for at least one input state $\rho_\S$ and $\sum_x p(x) \ketbra{x}{x}_\G$ the state $\sum_x q(x) \ketbra{x}{x}_\sC$ depends non-trivially on $\rho_\S$.

 {Thus $\Tr_{\G\S}(E_{s_\G}(\cdot)) : \S \to \sC$ is a non-trivial measurement of $\S$.}

 {This is a contradiction since we have created an interaction $E_{s_\G}(\cdot): \S \to \S \sC$ which is non-trivial non-disturbing measurement for an irreducible system.}

\subsection{The role of the no-signalling assumption}

 {Observe that 
\begin{align}
    (R^{-1} \otimes \I_\sC)\left(\sum_x q(x)\ketbra{x}{x} \otimes \sigma_\S(x) \otimes \ketbra{x}{x}_\sC\right) = \sum_x p(x) \ketbra{x}{x}_\G \otimes \rho_\S \otimes \ketbra{x}{x}_\sC . 
\end{align}
However the state of $\sC$ before the interaction was $\sum_x p(x) \ketbra{x}{x}_\sC$ whereas after the interaction it is $\sum_x q(x) \ketbra{x}{x}_\sC$. This is in general inconsistent with the fact that the action of $R^{-1} \otimes \I_\sC$ should leave $\sC$ unchanged.}

 The assumption that the action of $R^{-1} \otimes \I_\sC$ should leave $\sC$ unchanged is sometimes known as no-signalling, since it says that an operation on $\S\G$ should not affect $\sC$. It is met by quantum theory and is an assumption of the GPT framework more generally.
 
 If we impose no-signalling then  the interaction is only consistent when $p(x) = q(x)$ and hence 
\begin{align}
    R:: \sum_x p(x) \ketbra{x}{x} \otimes \rho_\S \mapsto \sum_x p(x) \ketbra{x}{x} \otimes \sigma_\S(x) ,
\end{align}
i.e. there is no back-reaction from $\S$ to $\G$. This is consistent with the main theorem.

\end{document}